\documentclass[twocolumn,fleqn]{article} 
\usepackage{amsmath,amssymb,amsfonts}
\usepackage[dvipdfmx]{graphicx} 	
\usepackage{url}
\usepackage{siunitx} 
\title {Radioactive contamination survey for 12 years in Kashiwa City, Chiba Prefecture, Japan}
\author {Yoshihiko Takase \thanks {Chiba JICA Senior Volunteers Association}} 

\date{}					

\begin{document} 
\maketitle 

\abstract {}
The Fukushima nuclear disaster on March 11, 2011 has been reported to the IAEA and has been dealt with professionally. 
On the other hand, airborne monitoring was conducted in September 2011 for wide-area air dose rate surveys. As a result, Kashiwa City in Chiba Prefecture, which is about $200 \mathrm{km}$ away, was designated as an "intensive contamination survey area." Localized radioactive contaminations (hotspots) were often found and reported by residents there. 
The present study is a report of 2011 and 2012 dose rate measurements and 12 years of fixed-point monitoring at hotspots in Kashiwa City.
A typical hotspot was a gutter and similar place where a large amount of rainwater was concentrated. The intensity distribution and time dependence of the dose rate at hotspots were measured. The dose rate over 12 years by fixed-point measurements agreed well with the sum of the radioactive decay characteristics of $^{131}\mathrm{I}$, $^{134}\mathrm{C_s}$, and $^{137}\mathrm{C_s}$. Decontamination in the living environment was conducted by sharing information between residents and the local government. 
It was effective to reduce the dose rate significantly by removing the sediment in the gutter and similar place. 
It is important to find hotspots in the living environments and decontaminate them intensively to reduce the risk of radiation injury to young children.

\section {Introduction}
\subsection {Recent evidences and original radiation data on the Fukushima nuclear disaster}

Twelve years have passed since the accident at the Fukushima Daiichi Nuclear Power Station (hereafter referred to as the Power Station), which was classified as an INES Level 7 Major accident\cite{INES}. At the time of the accident on March 11, 2011, Unit 1, Unit 2, and Unit 3 reactors were in operation. In Unit 1, which caused the first meltdown among them, 12 years after the accident (late March 2023), a robot camera was inserted under the reactor pressure vessel and the inside of the containment vessel was confirmed for the first time\cite{NHKscicul2023}.

The investigation inside the reactor containment vessel with a robot camera was long hampered by strong radiation and debris. In Unit 3, in July 2017, about 6 years after the accident, a robot that moved underwater was inserted under the pressure vessel to investigate the internal conditions. Deposits such as rock-like black masses, which were highly likely to be fuel debris, were confirmed. In Unit 2, in January 2018, same trial, and on February 13, 2019, a robot arm directly touched the debris\cite{NHKscicul2019}.

The accident at the Power Station was disclosed in the report of the Nuclear Emergency Response Headquarters (as of April 8, 2011)\cite{Report_NERH} and the report submitted to the IAEA by the Government of Japan (June 2011)\cite{Report_IAEA}. 

\begin {figure} 
	\centering 
	\includegraphics[width=8.0cm]{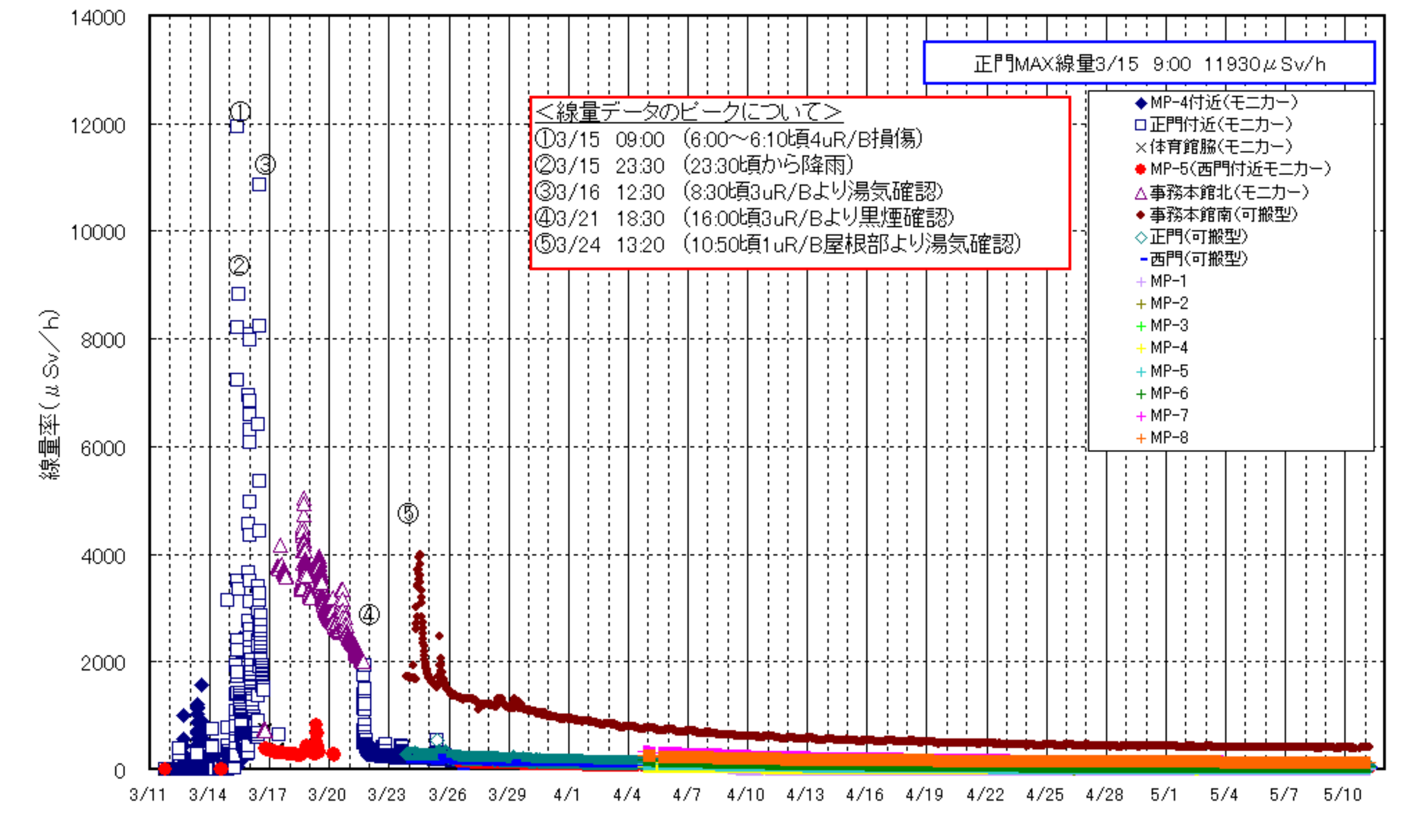} 
	\caption {The dose rate measurements at the Fukushima Daiichi V-9 monitoring post. \\
	Maximum dose rate at the main gate: $11930\mathrm{\mu{Sv/h}}$ at March 15 9:00.\\
	About dose data peaks \\
	(1) Mar-15 09:00 (around 6:00 - 6:10 Unit 4 reactor building damage) \\
	(2) Mar-15 23:30 (Rain from around 23:3) \\
	(3) Mar-16 12:30 (Steam confirmed from Unit 3 reactor building around 8:30) \\
	(4) Mar-21 18:30 (Graphite confirmed from Unit 3 reactor building around 16:00) \\
	(5) Mar-24 13:20 (Steam confirmed from Unit 1 reactor building roof around 10:50) \\
	}
	\label {fig:dosepeak_p162} 
\end {figure}

Figure \ref{fig:dosepeak_p162} introduces the dose rate measurements at the Fukushima Daiichi V-9 monitoring post\cite{Report_IAEA} from among the vast amount of data. This included the detection of $11.93\mathrm{ mSv/h}$ of gamma rays, the highest ever recorded, near the main gate at 09:00 on March 15, as well as the dates and associated events of the five peak dose data.

\subsection {Wide area airborne monitoring}

The Ministry of Education, Culture, Sports, Science and Technology has conducted airborne monitoring within a range of $100 \mathrm{km}$ from the  Power Station, as well as Saitama and Chiba Prefectures, in order to understand the effects of radioactive materials over a wide area\cite{Aircraft_monitoring}.

\begin {figure} 
	\centering 
	\includegraphics[height=7.0cm]{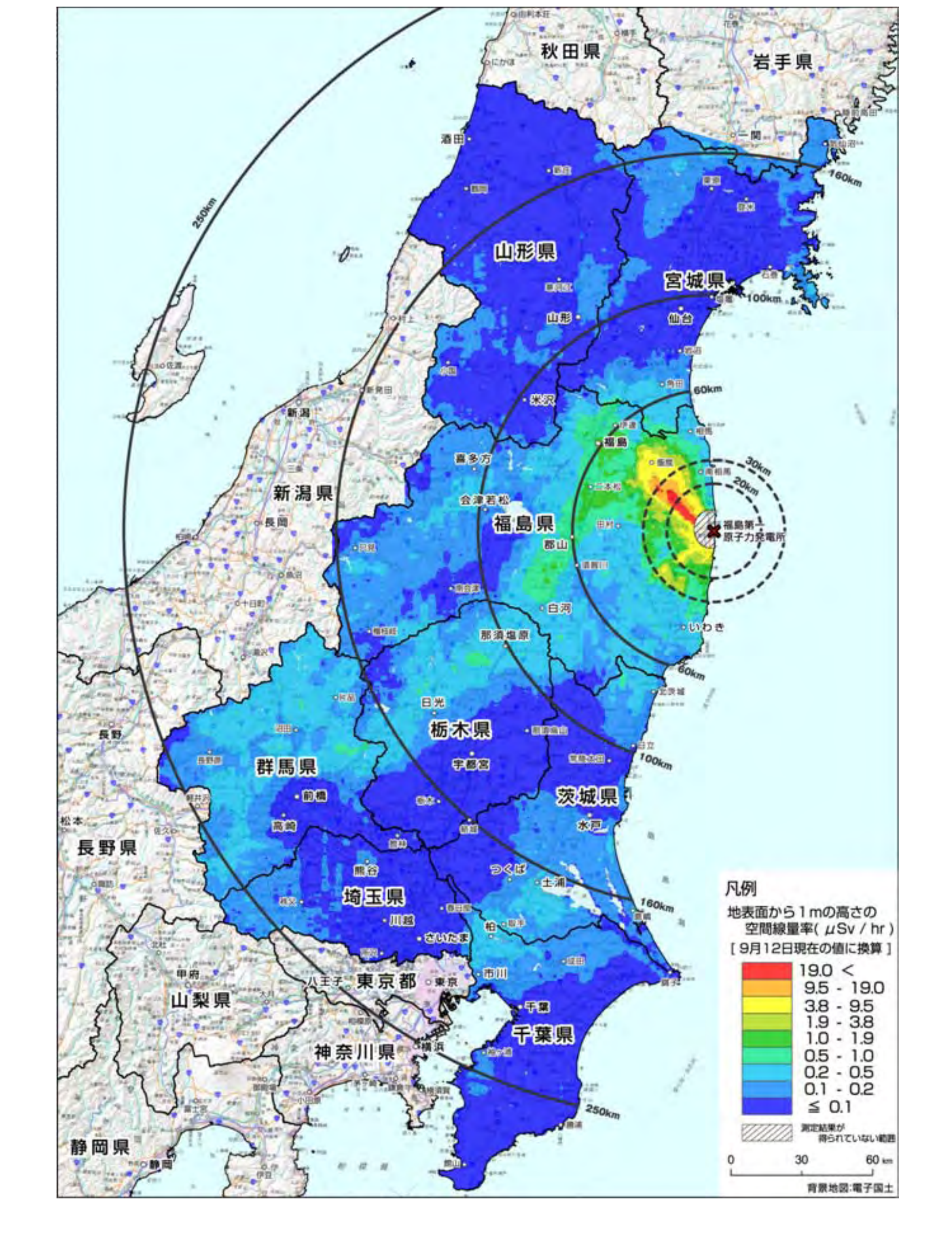} 
	\caption {The air dose rate at a height of $1 \mathrm{m}$ above the ground surface (converted to values as of September 12, 2011) measured by airborne monitoring conducted by the Ministry of Education, Culture, Sports, Science and Technology.
	}
	\label {fig:dose_survay250km} 
\end {figure}

As an example, Fig. \ref{fig:dose_survay250km} shows the air dose rate measurement results at a height of $1\mathrm{m}$ above the ground ($h=\mathrm{1m}$).

The highest air dose rate was measured in the northwest direction from the Power Station. In addition, relatively high dose rates were measured in Kashiwa City, Chiba Prefecture, and Ibaragi Prefecture, about $200 \mathrm{km}$ south-southwest from the Power Station. As a countermeasure to the latter, such a area was designated as an "intensive contamination survey area", and if a place where the air dose rate at a height of $1\mathrm{m}$ above the ground was $1 \mathrm{{\mu}Sv/h}$ or more higher than the surrounding area was discovered, it was decided to carry out decontamination (removing mud from gutters, collecting fallen leaves, pruning trees, washing with water, brushing, etc.)\cite{Guideline-localcontami}. In fact, measures such as investigation and decontamination were examined at locally contaminated spots (hereafter referred to as hotspots) found in Kashiwa City\cite{Kashiwa-contami_survay}.

According to the commentary by Yamazawa et al.\cite{Yamazawa-Hirao}, it was estimated that the emission rate of radioactive materials on the 15th was one to two orders of magnitude higher than on other days. Transport to the inland by the sea breeze and valley breeze that occurred at this time and retention by the weak wind at night was one of the factors that form the polluted area in terms of atmospheric transport. Furthermore, precipitation was observed from the evening of the same day to the next day. It was concluded that the northwestward contaminated area was caused by the combination of three conditions: the high emission rate on March 15, the local wind circulation, and the precipitation.

In Kashiwa City, radioactive $\mathrm{C_s}$ was detected in incinerated ash (60.8 to 70.8 kBq from solidified ash, far exceeding the government-decided landfill limit of 8 kBq) at an incineration plant, which had a serious impact on waste disposal.
Hotspots were often found and reported by residents there. Twelve years later, many of these reports have disappeared from the website as well. It seems important to record at least some of them.

The present study is a report of 2011 and 2012 dose rate measurements and 12 years of fixed-point monitoring at hotspots in Kashiwa City.
An electronic dosimeter DoseRAE 2 was used after confirming its reliability experimentally\cite{DoseRAE2test}.

\section {Dose rate measurement at hotspots in Kashiwa City} 
\subsection {Parking lot gutter}

\begin {figure} [p] 
	\centering 
	\includegraphics[width=6.0cm]{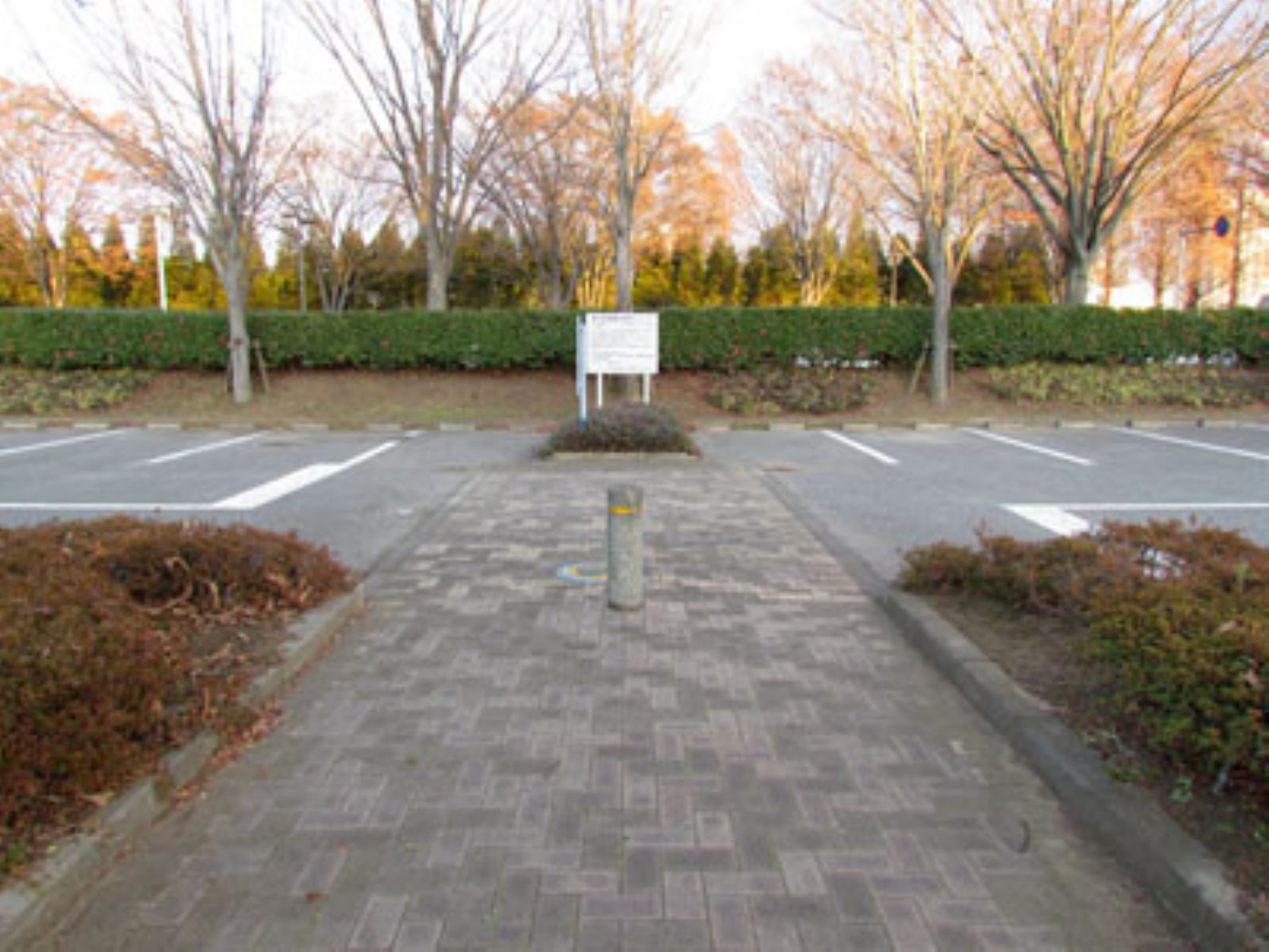} 
	\caption {Photo of a parking lot of Kashiwanoha Park with a sidewalk in the middle and gutters at the back.
	}
	\label {fig:parking} 
\end {figure}

\begin {figure} [p] 
	\centering 
	\includegraphics[width=8.0cm]{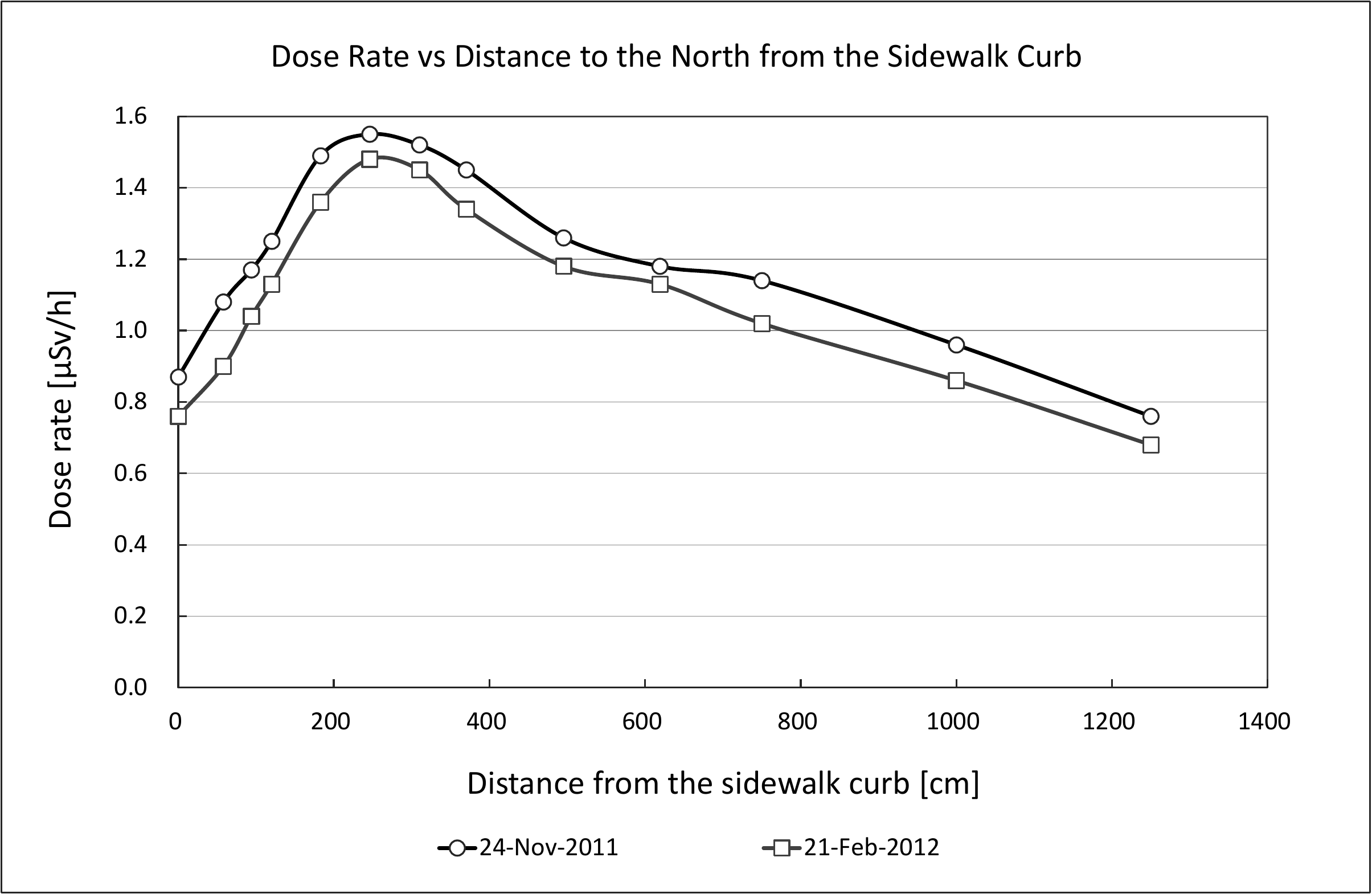} 
	\caption {The relationship between the dose rate and the distance to the north (to the left in Fig. \ref{fig:parking}).
	}
	\label {fig:dose_x_parkingnorth} 
\end {figure}

\begin {figure} [p] 
	\centering 
	\includegraphics[width=8.0cm]{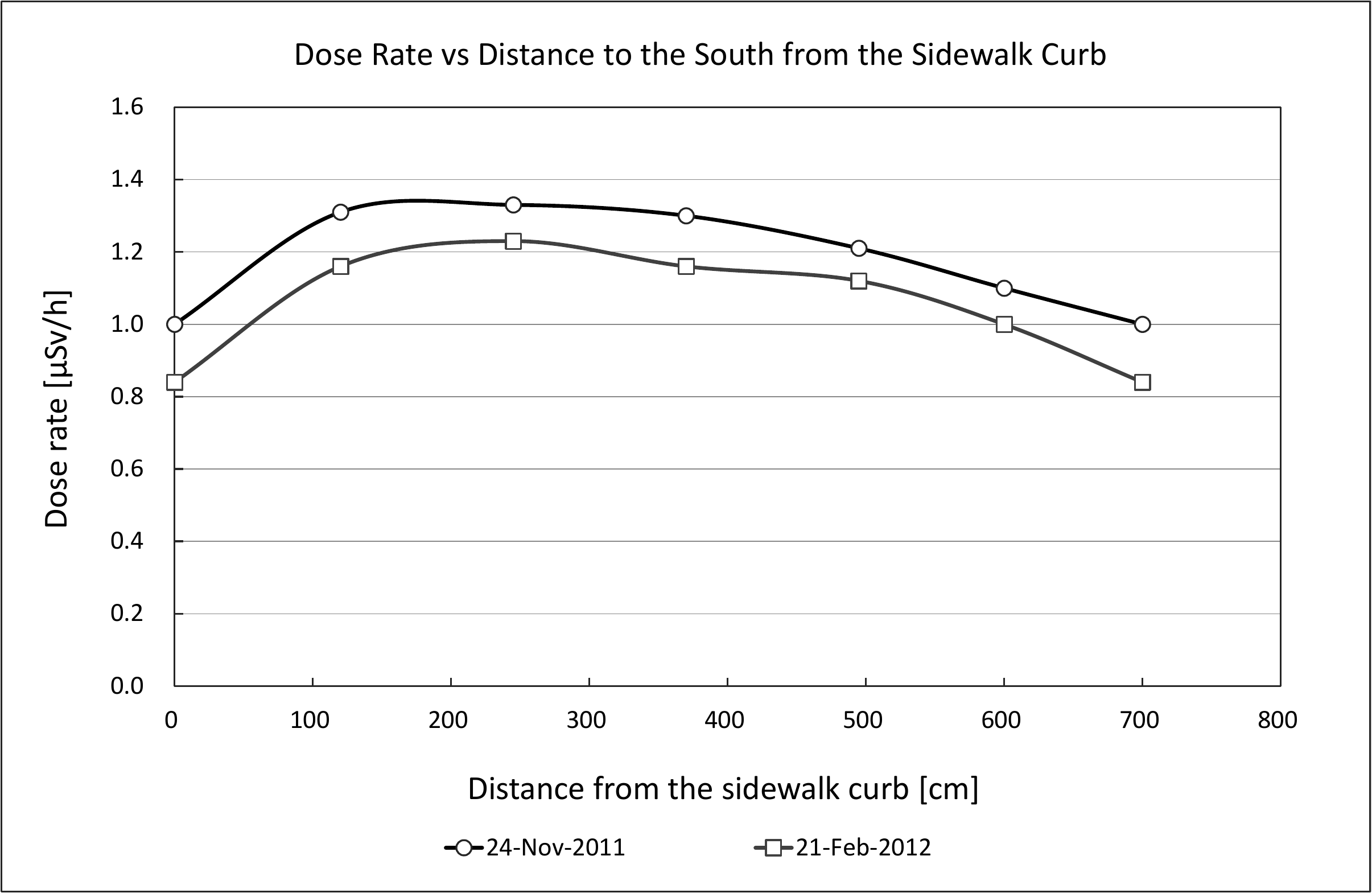} 
	\caption {The relationship between the dose rate and the distance to the south (to the right in Fig. \ref{fig:parking}).
	}
	\label {fig:dose_x_parkingsouth} 
\end {figure}

\begin {figure} [p] 
	\centering 
	\includegraphics[width=8.0cm]{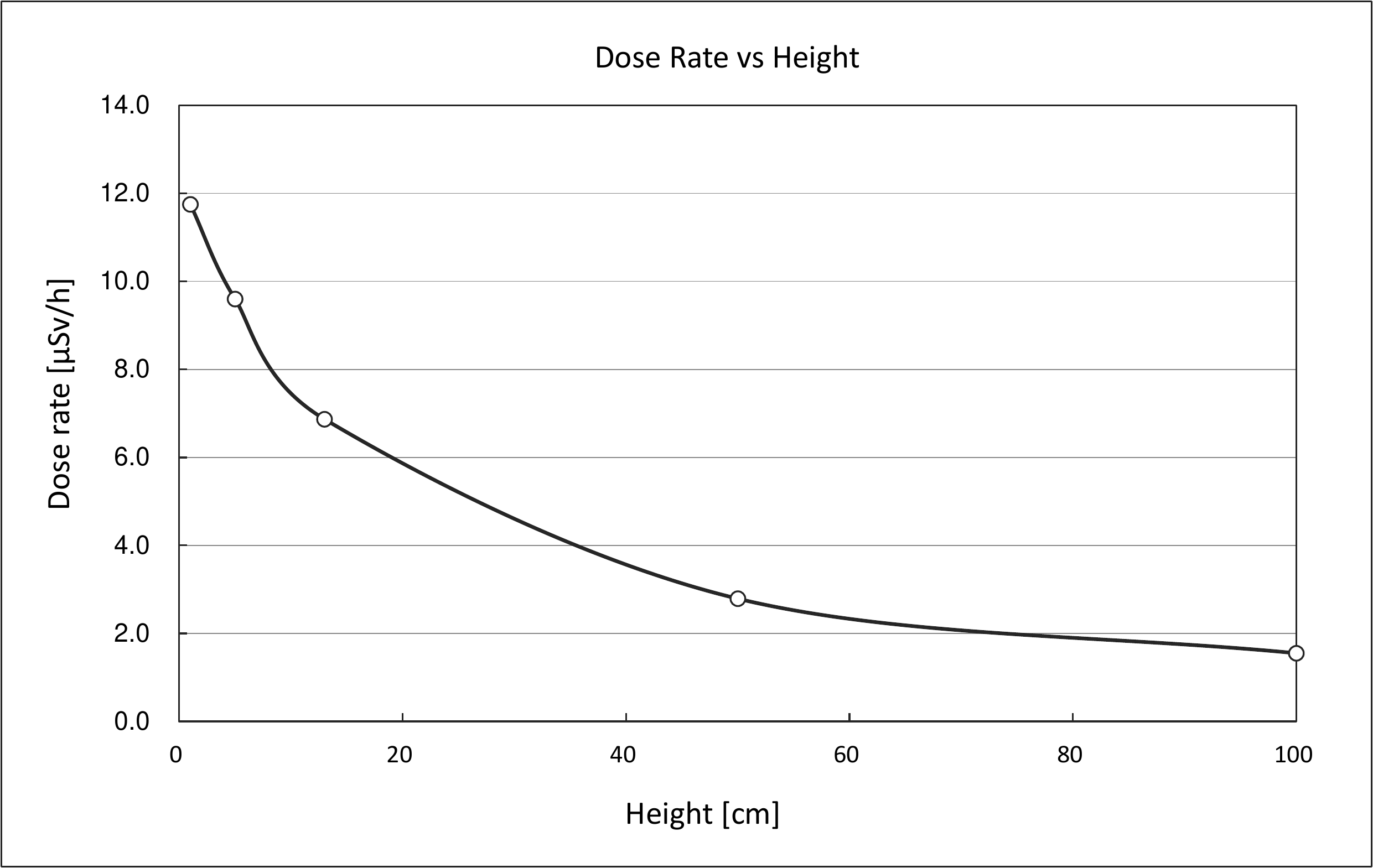} 
	\caption {The relationship between the dose rate and the height.
	}
	\label {fig:dose_h_parking} 
\end {figure}

Figure \ref{fig:parking} shows a photo of a parking lot of Kashiwanoha Park with a sidewalk in the middle and gutters at the back (area$=2\times\mathrm{700m^2}$).
Figures \ref{fig:dose_x_parkingnorth} and \ref{fig:dose_x_parkingsouth} show the relationship between the dose rate and the distance to the left (north) and right (south) in the photo, respectively. The dose rate at a height of $1\mathrm{m}$ exceeded $1 \mathrm{{\mu}Sv/h}$ over about $9 \mathrm{m}$ in the northern direction, and the maximum value was $1.55 \mathrm{{\mu}Sv/h}$.  Figure \ref{fig:dose_h_parking} shows the dose rate dependence on height. It was $11.75 \mathrm{{\mu}Sv/h}$ ($h\approx{1\mathrm{cm}}$) and $1.55 \mathrm{{\mu}Sv/h}$ ($h={1\mathrm{m}}$).

\begin {figure} [p] 
	\centering 
	\includegraphics[width=6.0cm]{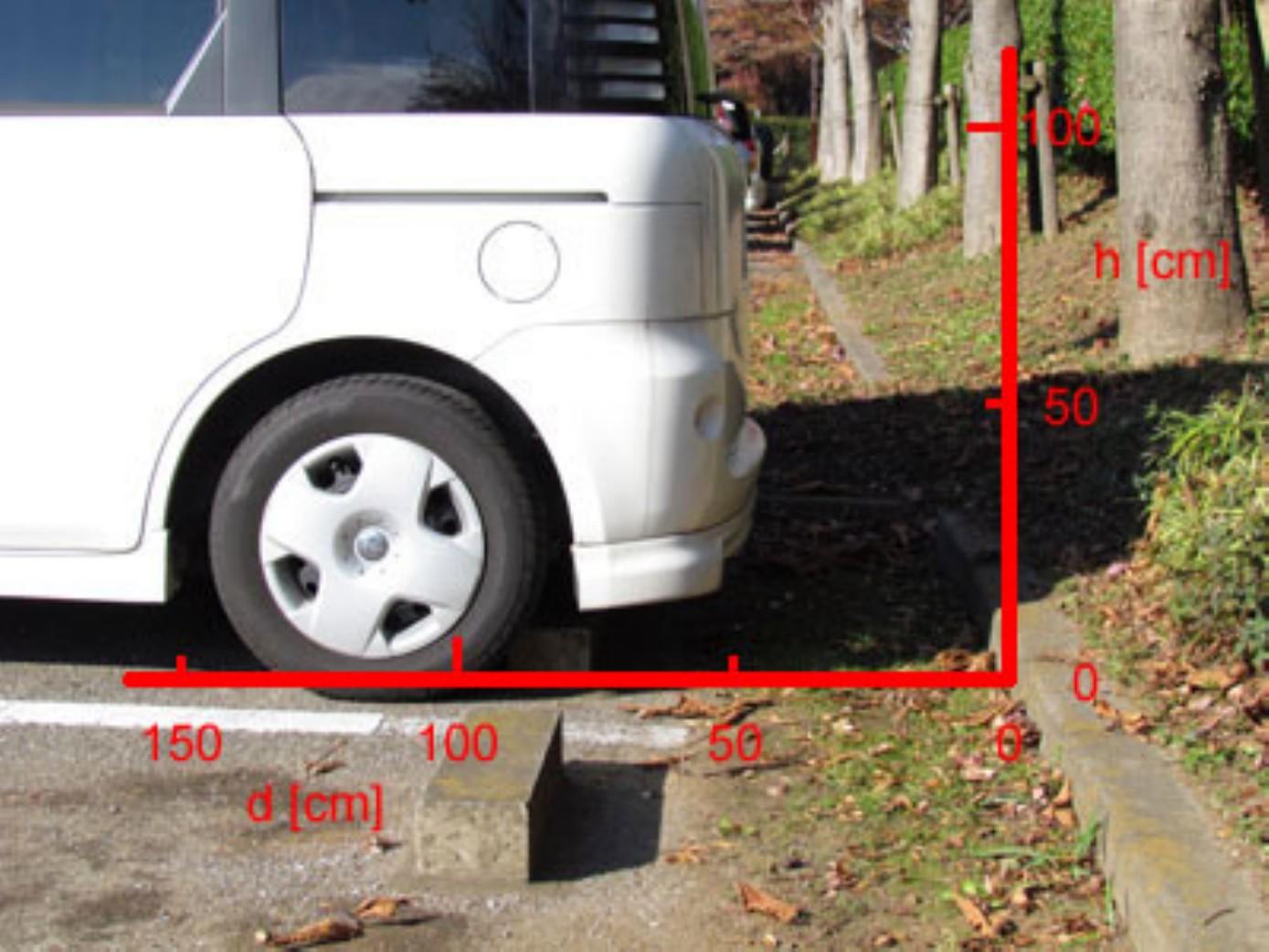} 
	\caption {Photograph of parking lot with embankment curb on the right.
	}
	\label {fig:parking2west} 
\end {figure}

\begin {figure} [p] 
	\centering 
	\includegraphics[width=8.0cm]{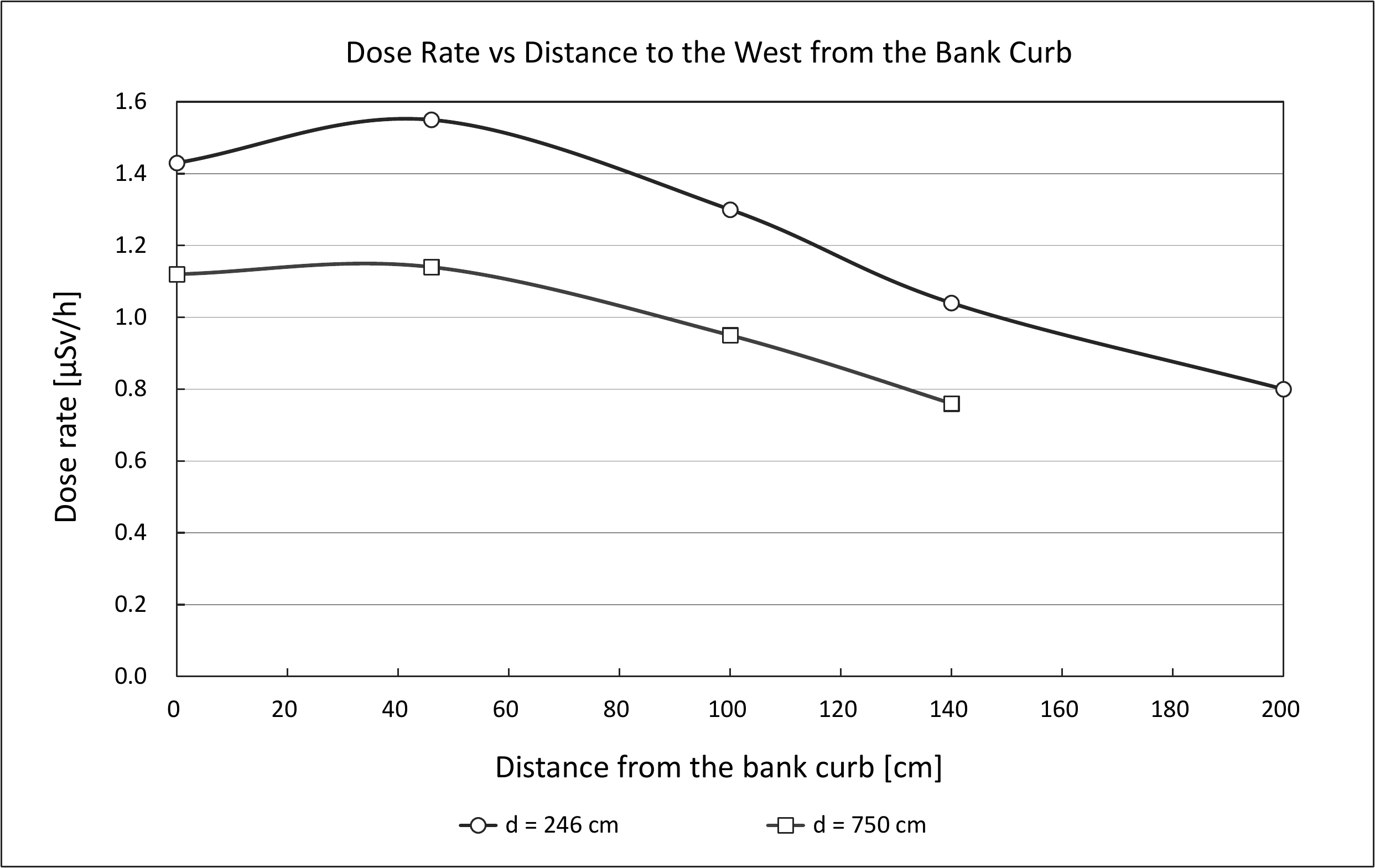} 
	\caption {The relationship between the dose rate and the distance to the west (to the left in Fig. \ref{fig:parking2west}).
	}
	\label {fig:dose_x_parkingwest} 
\end {figure}

Figure \ref{fig:parking2west} is a photo of the parking lot with the embankment curb on the right.
Figure \ref{fig:dose_x_parkingwest} shows the relationship between the dose rate and the distance in the west direction (to the left of the photo, 246 and 750cm north of the sidewalk curb in Fig. \ref{fig:parking}). In areas where the dose rate was high, the dose rate exceeded $1 \mathrm{{\mu}Sv/h}$ up to $\mathrm{150 cm}$ from the embankment curb and the maximum value was $1.55 \mathrm{{\mu}Sv/h}$ ($h={1\mathrm{m}}$).

The earth and sand accumulated in the gutter did not flow out because short grasses were rooted in it. For this reason, the radioactive $\mathrm{C_s}$ contained in the rain that fell on the large parking lot was thought to have been adsorbed by the sediment and became a stable and highly radioactive contamination source. This is expressed here as the "gutter effect".

\subsection {Paved road gutter}

\begin {figure} [p] 
	\centering 
	\includegraphics[width=6.0cm]{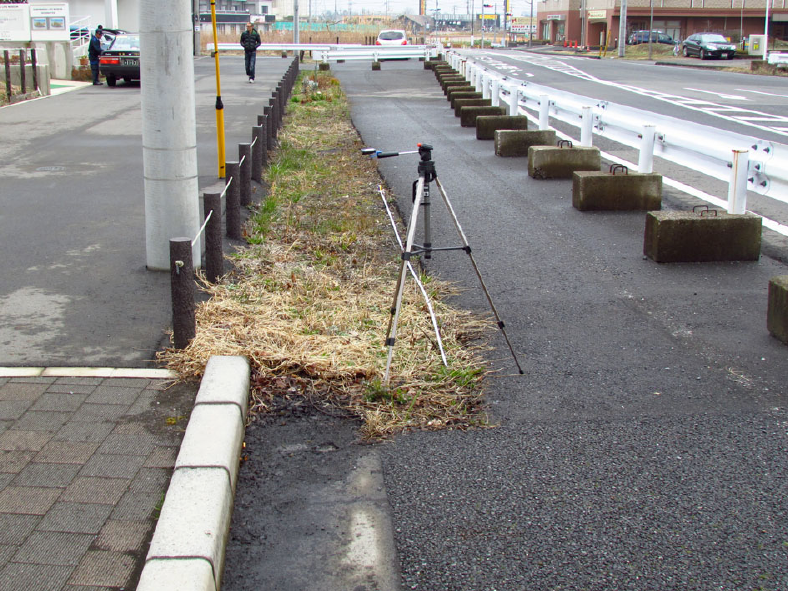} 
	\caption {Photo of shrubbery at the edge of a paved wide intersection.
	}
	\label {fig:paved-road-gutter} 
\end {figure}

Another example of a familiar hotspot was a drainage channel where radioactive rainfall that fell on a large paved area collected. This is also the "gutter effect". As an example, Fig. \ref{fig:paved-road-gutter} shows a place where part of the rain that fell on a wide paved intersection flew into the shrubbery at the edge of the intersection.

\begin {figure} [p] 
	\centering 
	\includegraphics[width=8.0cm]{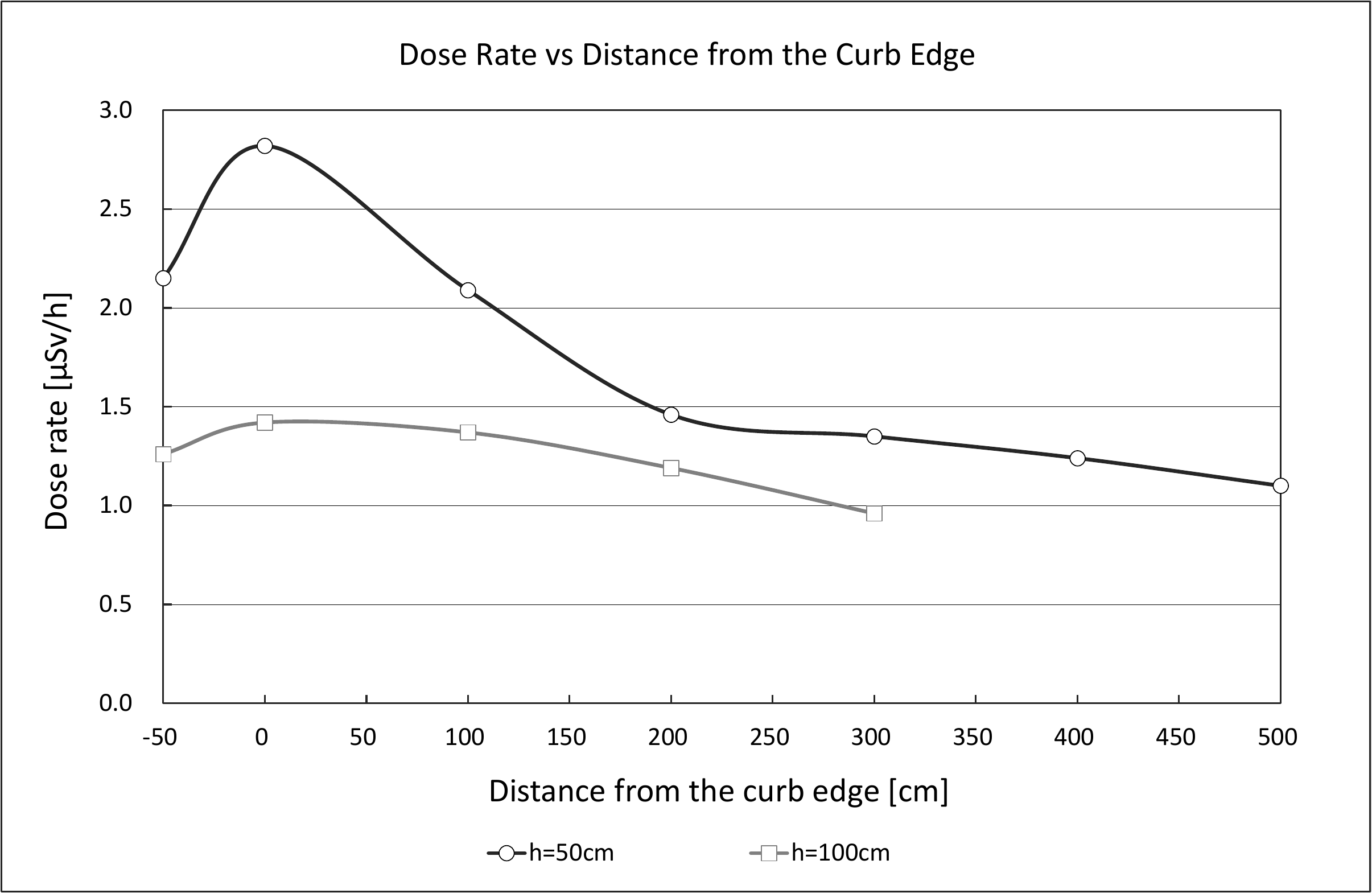} 
	\caption {The distance dependence of the dose rate at heights of $50 \mathrm{cm}$ and $100 \mathrm{cm}$.
	}
	\label {fig:dose-x_paved-road-gutter} 
\end {figure}

Figure \ref{fig:dose-x_paved-road-gutter} shows the distance dependence of the dose rates at heights of $50 \mathrm{cm}$ and $100 \mathrm{cm}$. The distance here refers to the distance measured in the back direction from the white curb in front of the sidewalk seen in the photo. 
The highest dose rates were $2.82 \mathrm{\mu{Sv/h}}$ ($h={50\mathrm{cm}}$) and $1.42 \mathrm{\mu{Sv/h}}$ ($h={100\mathrm{cm}}$) at $0 \mathrm{cm}$ distance, where the dose rate at $1\mathrm{m}$ height exceeded $1 \mathrm{\mu{Sv/h}}$ over $2 \mathrm{m}$ distance.

\begin {figure} [p] 
	\centering 
	\includegraphics[width=8.0cm]{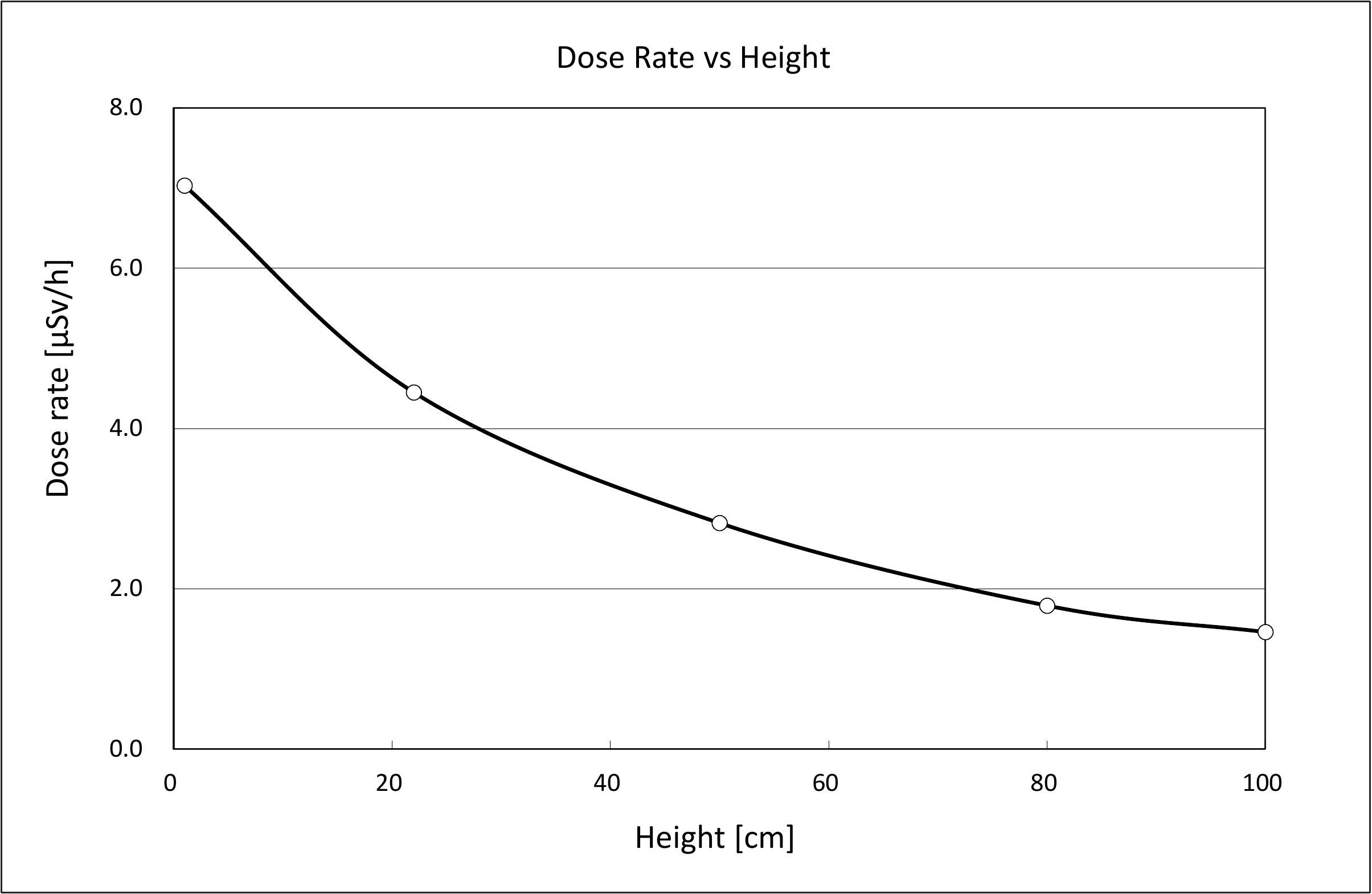} 
	\caption {The height dependence of the dose rate at the place where it was highest.
	}
	\label {fig:dose-h_paved-road-gutter} 
\end {figure}

Figure \ref{fig:dose-h_paved-road-gutter} shows a graph of the measured dose rate as a function of height at the point where the dose rate was highest. Near the surface of the soil ($h\approx{1\mathrm{cm}}$), the value was as high as $7.03 \mathrm{{\mu}Sv/h}$.

\subsection {Bases of large zelkova trees}

It was also found that the bases of large zelkova trees, which were roadside trees, became the hotspots. 
This is also the "gutter effect". When we measured and compared the bases of various trees, the "gutter effect" of the zelkova tree was especially remarkable.

\begin {figure} [p] 
	\centering 
	\includegraphics[height=8.0cm]{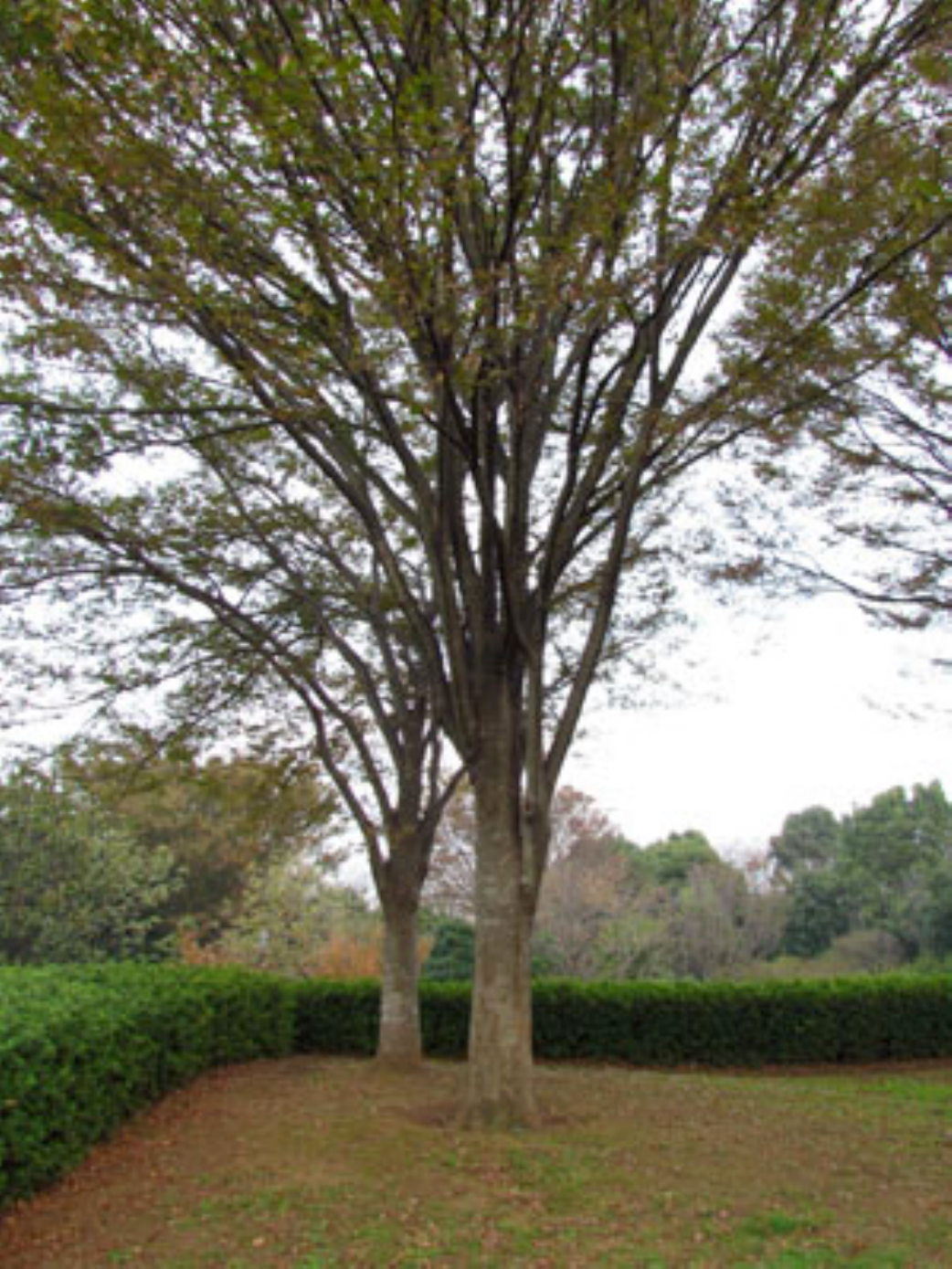} 
	\caption {Photo of a large zelkova tree.
	}
	\label {fig:zelkova} 
\end {figure}

\begin {figure} [p] 
	\centering 
	\includegraphics[width=8.0cm]{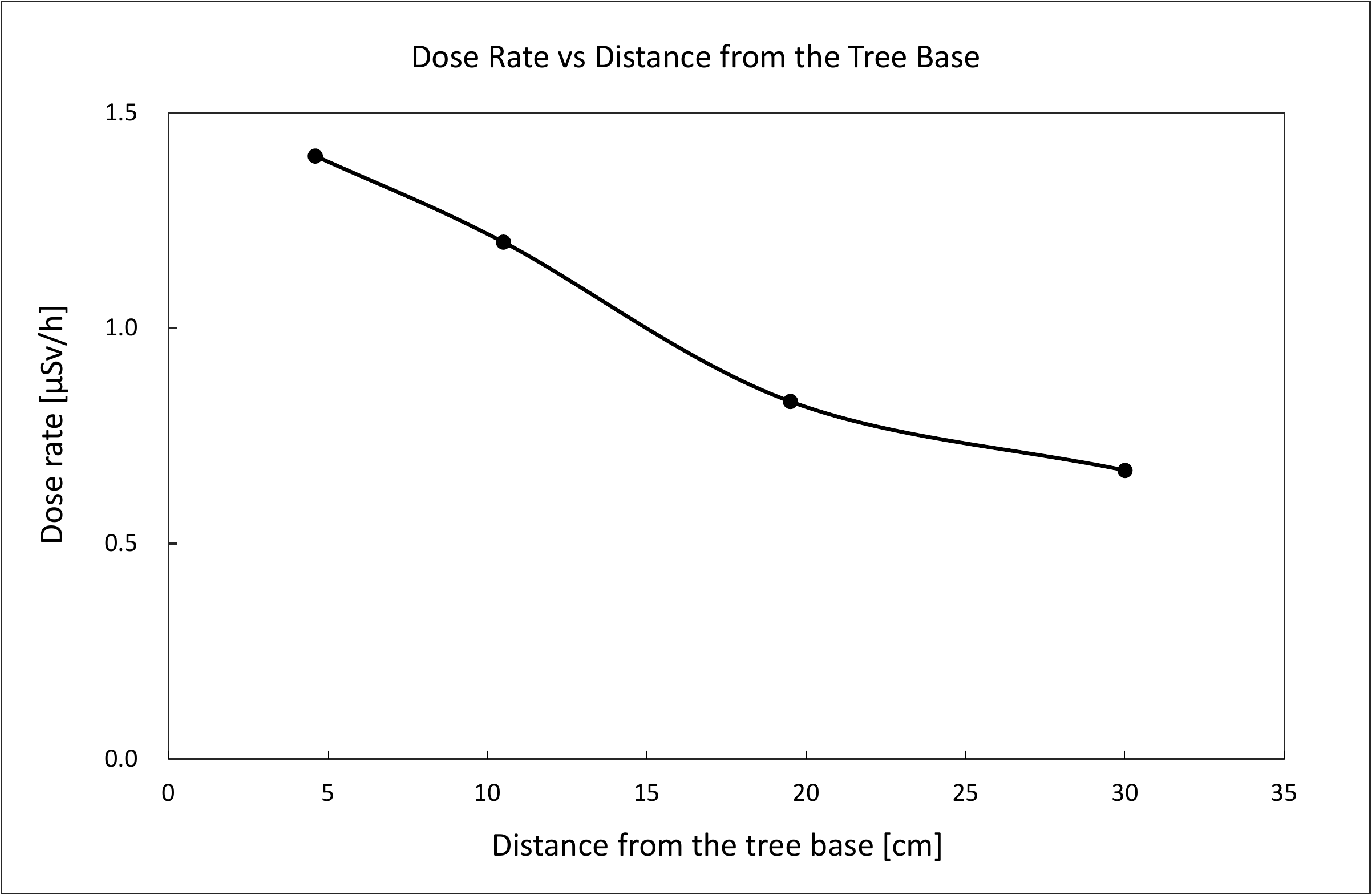} 
	\caption {The dose rate depending on the distance from the base of the tree ($h\approx{1\mathrm{cm}}$).
	}
	\label {fig:dose_x_treebase} 
\end {figure}

Figure \ref{fig:zelkova} shows a photo of a large zelkova tree and Fig. \ref{fig:dose_x_treebase} a graph showing an example of changes in radiation dose rate depending on the distance from the base of the tree ($h\approx{1\mathrm{cm}}$).

Table \ref{table:doses-zelkovatrees} shows the dose rates measured at the bases of four zelkova trees in Kashiwanoha Park.

\begin{table} 
  \caption {The dose rates [$\mathrm{{\mu}Sv/h}$] at the bases of four zelkova trees ($h\approx{1\mathrm{cm}}$) in Kashiwanoha Park.\\} 
  \label {table:doses-zelkovatrees}
  \centering 
  \begin {tabular} {crrrr} 
	\hline 
	Place & \multicolumn{1}{c}{Site A} & \multicolumn{1}{c}{Site B} & \multicolumn{1}{c}{Site C} & \multicolumn{1}{c}{Site D} \\ 
	\hline \hline 
	Dose rate & 3.51 & 3.08 & 3.21 & 2.36 \\ 
	\hline
  \end {tabular} 
\end {table}

Around March, when the Power Station accident occurred, the zelkova trees were spreading upside-down bamboo broom-like branches without growing leaves. Since the zelkova bark is smooth, it was thought that rainwater was efficiently collected from the branches along the trunk to the roots. Since zelkova is often planted in places where people gather, it was imagined that children often touch the bases.

\section {Environmental dose decrease} 
\subsection {Decontamination}

From February 10 to 16, 2012, Chiba Prefecture measured radioactive contamination in Kashiwanoha Park, and prohibited entry to areas showing high radiation levels. On February 16, it was announced that preparations were underway for immediate decontamination of the affected areas.

\begin {figure} 
	\centering 
	\includegraphics[width=7.0cm]{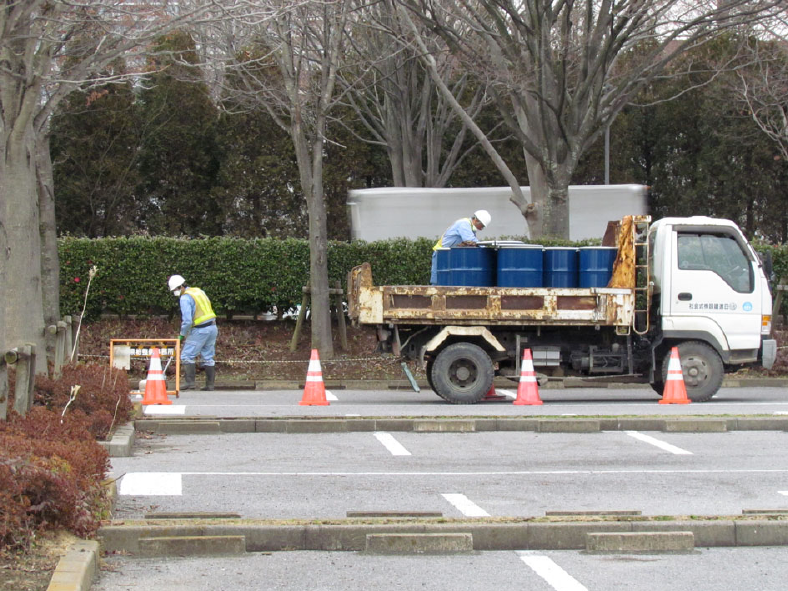} 
	\caption {Photo of decontamination work in the parking lot.
	}
	\label {fig:decontami_parking} 
\end {figure}

\begin {figure}
	\centering 
	\includegraphics[width=7.0cm]{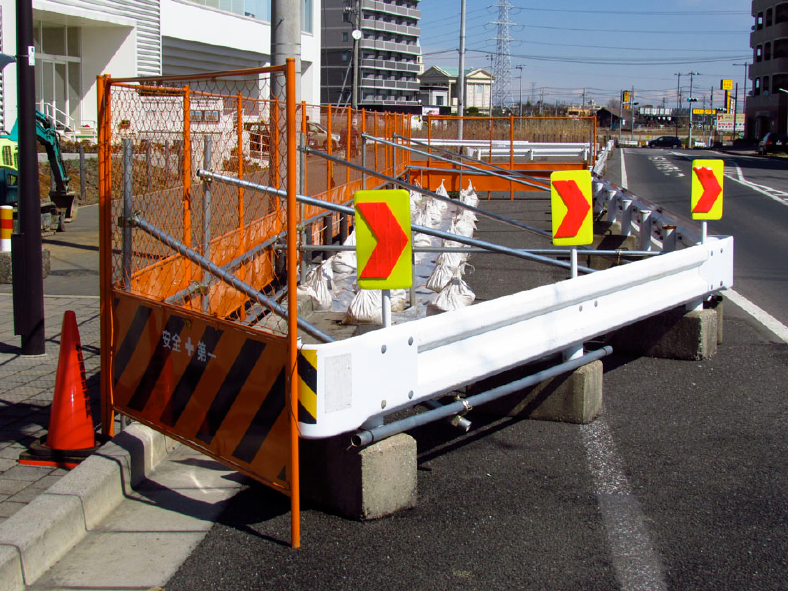} 
	\caption {Photo of decontamination work at the same place in Fig. \ref{fig:paved-road-gutter}.
	}
	\label {fig:decontami_crossing} 
\end {figure}

The parking lot was decontaminated on March 7, 2012. The work consisted of removing the highly radioactive sediment, packing it in drums and moving them to a temporary storage site (Fig. \ref{fig:decontami_parking}). Decontamination work at the same place in Fig. \ref{fig:paved-road-gutter} was also carried out as shown in Fig. \ref{fig:decontami_crossing}.

At the point where the dose rate was $11.75 \mathrm{{\mu}Sv/h}$ ($h\approx{1\mathrm{cm}}$) before decontamination (November 24, 2011), it decreased significantly to $0.67 \mathrm{{\mu}Sv/h}$ after decontamination (March 10, 2012). At one of other points, it also decreased significantly from $10.42 \mathrm{{\mu}Sv/h}$ ($h\approx{1\mathrm{cm}}$) to $0.60 \mathrm{{\mu}Sv/h}$. It may be concluded that the decontamination work was carried out effectively, although long-term management is required to store the removed radioactively contaminated sediment.

\subsection {Dose rate monitoring in the daily living environment}

\begin {figure} 
	\centering 
	\includegraphics[width=8.0cm]{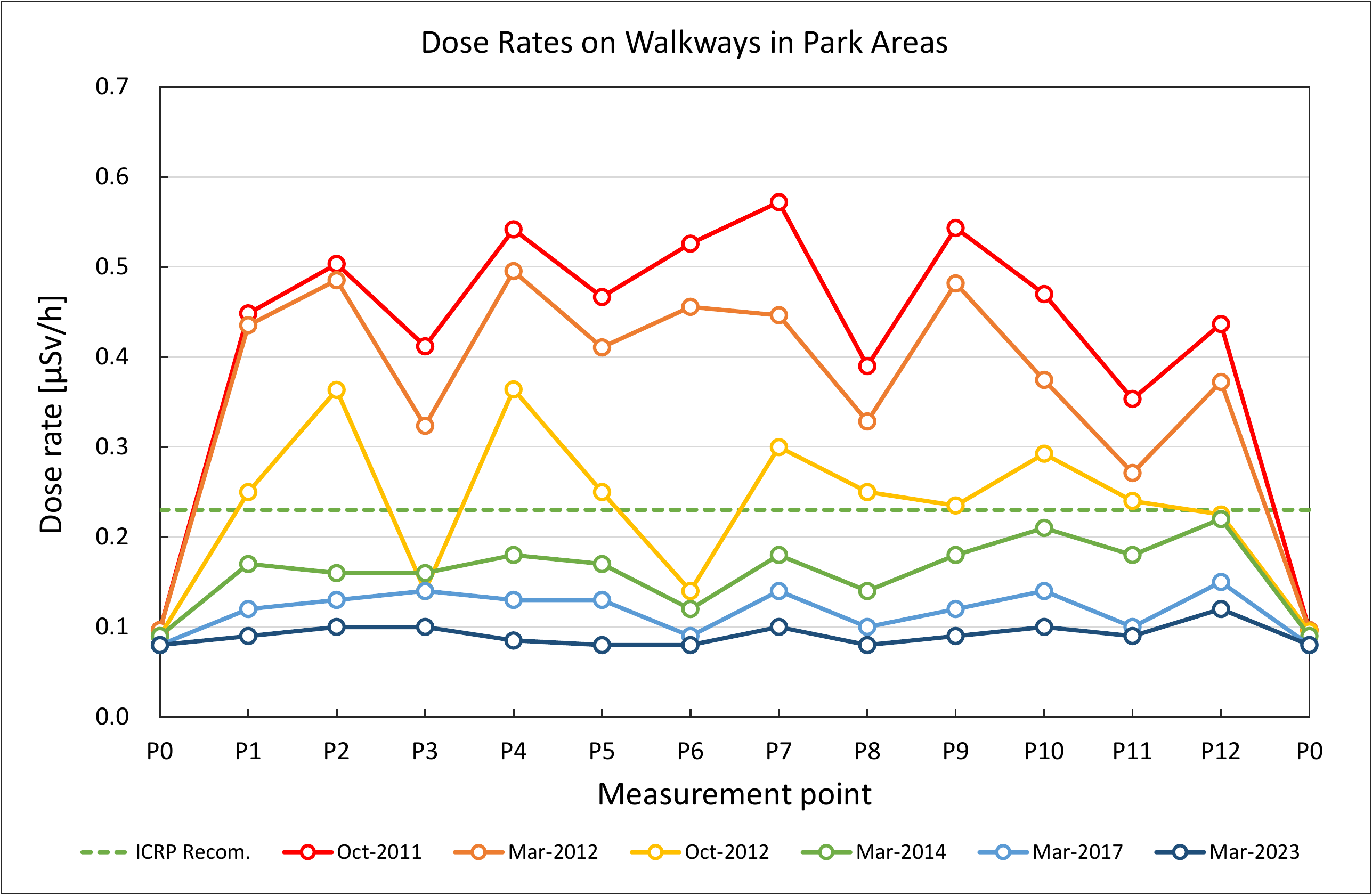} 
	\caption {The time dependence of the dose rates on walkways in Kashiwanoha Park area for 12 years. P0: inside the house, P1: sidewalk1 ($h=1\mathrm{m}$), P2: sidewalk2 ($h=1\mathrm{m}$), P3: on the park stone gate, P4: on the park outdoor stage, P5: park walking path1 ($h\approx{70\mathrm{cm}}$), P6: park lawn plaza ($h\approx{1\mathrm{cm}}$), P7: park walking path2 ($h\approx{70\mathrm{cm}}$), P8: park walking path3 ($h\approx{70\mathrm{cm}}$), P9: park sakura plaza ($h\approx{70\mathrm{cm}}$), P10: park cobbled sidewalk ($h=1\mathrm{m}$), P11: main gate of a high school ($h=1\mathrm{m}$), and P12: on the flowerbed curb in front of the house.
	}
	\label {fig:dose_place_walking} 
\end {figure}

In order to know the dose rate in the daily living environment, we have been measuring the time dependence of the dose rates on walkways in Kashiwanoha Park area for 12 years. Those measurement data (October 2011, March 2012, October 2012, March 2014, March 2017, and March 2023) are shown in Fig. \ref{fig:dose_place_walking}.

As can be seen in Fig. \ref{fig:dose_place_walking}, the radiation dose rate has decreased after three years to values below the International Commission on Radiological Protection (ICRP) recommendation ($0.23 \mathrm{{\mu}Sv/h}$)\cite{ICRP}. Locations with significant dose rate drops (P3, P6, and P9) are decontaminated places within the park.

\section {Discussions} 
\subsection {Rutherford and Soddy law}

During early studies of radioactive materials, Rutherford and Soddy\cite{Rutherford} found that the rate of the number of nuclei that decay radioactively per unit time is constant, and that radioactivity follows an exponential decay law with time. In other words, the number of nuclei decaying per unit time is proportional to the number of nuclei remaining at that time, and can be expressed by the following differential equation.

\begin {equation} 
\label {eq:rutherford} 
  -\frac{\mathrm{d}N(t)}{\mathrm{d}t} = {\lambda}N(t), 
\end {equation} 

where $N(t)$ is the number of remaining nuclei at time $t$, $\lambda$ is the decay constant specific to the nuclide (reciprocal of average lifetime), and the left side is the number of nuclei decaying per unit time (= radioactivity). The solution of this equation is the following equation.

\begin {equation} 
\label {eq:explaw} 
  N(t) = {N_0}e^{-{\lambda}t},
\end {equation} 

where $N_0$ is the number of nuclei at time $t = 0$. 
Radioactive decay is often expressed in half-lives $T$. The relational expression between the half-life and the decay constant is as follows.

\begin {equation} 
\label {eq:half-life} 
 \lambda{T} = \mathrm{ln}{2}.
\end {equation} 

Since the measured radiation dose rate, $Doserate$, is proportional to the radioactivity, we get

\begin {equation} 
\label {eq:doserate} 
  Doserate = \lambda{N_0}e^{-{\lambda}t}.
\end {equation} 

If we take the natural logarithm of both sides,

\begin {equation} 
\label {eq:lndoserate} 
  \mathrm{ln}(Doserate) = - \lambda{t} + \mathrm{ln}{(\lambda{N_0})}. 
\end {equation} 

Therefore, the graph of $\mathrm{ln}(Doserate)$ and $t$ becomes a straight line with a slope of $-\lambda$.

\subsection {Reliability test of DoseRAE 2}

Experimental results for confirming the reliability of DoseRAE 2 using the radioactively contaminated spots as radiation sources were reported separately\cite{DoseRAE2test}. The main results are presented in the appendix. The conclusions were as follows.

When the dose rate $\approx\mathrm{0.1{\mu}Sv/h}$ the time for steady state $\approx\mathrm{10 min}$ and the display was stable. The deviations of measured values were within $\pm{15 \%}$ and the resolution was $0.01 \mathrm{{\mu}Sv/h}$.

When the dose rate $\approx\mathrm{0.5{\mu}Sv/h}$ the displayed value fluctuated greatly and the deviations of metastable measured values in about 2 minutes were within $\pm{30\%}$. However, reading metastability measurements five times in about 10 minutes and averaging them yielded deviations within $\pm{3 \%}$. The deviations of measured values were within $\pm{3 \%}$ when the average of all data was taken by video recording method.

When the dose rate $\approx\mathrm{1{\mu}Sv/h}$ the time for steady state $\approx\mathrm{3 min}$ and the display was stable. The deviations of measured values were within $\pm{3\%}$. 

When the dose rate $\approx\mathrm{10{\mu}Sv/h}$ the time for steady state $\approx\mathrm{3 min}$ and the display was stable. The deviations of measured values were within $\pm{3\%}$.

\subsection {The environmental radiation information of the Univ. of Tokyo}

The environmental radiation information of the University of Tokyo, which has a campus in Kashiwa City, is open to the public\cite{TokyoUnivERI}.

\begin {figure} 
	\centering 
	\includegraphics[width=8.0cm]{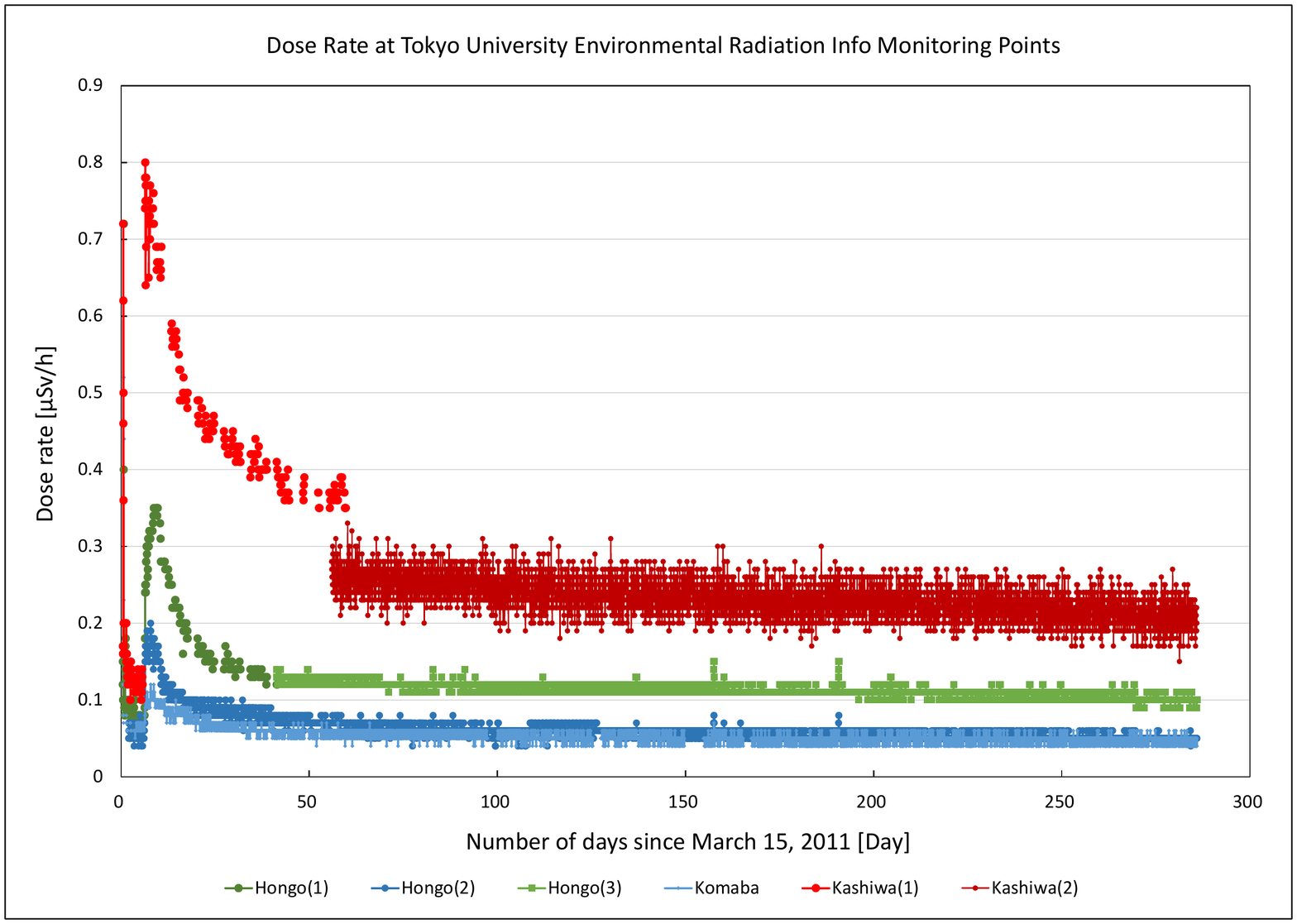} 
	\caption {The time dependence of the observed dose rate at 6 monitoring points (Hongo(1), Hongo(2), Hongo(3), Komaba(1), Kashiwa(1), and Kashiwa(2)) of the environmental radiation information of the University of Tokyo.
	}
	\label {fig:dose_t_tokyouniv} 
\end {figure}

Figure \ref{fig:dose_t_tokyouniv} shows the time dependence of observed dose rates at 6 monitoring points. Observations began on March 15, 2011 and ended on December 25, 2011. The data for Kashiwa City are Kashiwa(1) from March 15 to May 13 and Kashiwa(2) from May 10 to December 25.

The Kashiwa(1) dose rate, which was $0.16 \mathrm{\mu{Sv/h}}$ at 14:00 on March 15, jumped to $0.70 \mathrm{\mu{Sv/h}}$ at 14:30, and returned to the original value at 17:30. After that, the dose rate, which was about $0.12 \mathrm{\mu{Sv/h}}$ until March 20, increased sharply to about $0.75 \mathrm{\mu{Sv/h}}$ on March 21, and then continued to decrease slowly.

At 09:00 on March 15, 2011, the highest dose rate of $11.93 \mathrm{mSv/h}$ was measured near the main gate of the Power Station. This corresponds to 6:00, March 15, 2011 Unit 4 reactor building damage by hydrogen explosion as shown in Fig. \ref{fig:dosepeak_p162}.

Assuming that the radioactive plume traveled from the explosion at 6:00 on March 15 to the sharp peak at 14:30 on March 15 seen in Fig. \ref{fig:dose_t_tokyouniv}, the wind speed would be 6.4 m/s. However, no such wind was observed in the weather information for that day.

On the other hand, from around 4:00 to 7:00, it was recorded that a high radiation dose rate of 1 to $5.5 \mathrm{\mu{Sv/h}}$ was observed in Ibaragi Prefecture, gradually moving southward. As pointed out by Yamazawa et al.\cite{Yamazawa-Hirao}, the plume that affected the Kanto region on the 15th was emitted from midnight the previous day to the morning of the day, and it was possible that it was different from the plume that affected the northwest direction. Since there was no rainfall on that day, it was thought that the dose rate dropped sharply as the atmosphere passed.

From around 4:00 to 5:30 on March 21, high radiation dose rate was observed moving southward from northeastern to eastern Ibaragi Prefecture, with a maximum value of $2.908 \mathrm{\mu{Sv/h}}$ being observed in Hokota City. As pointed out by Yamazawa et al.\cite{Yamazawa-Hirao}, from the night of the 21st to the dawn of the 22nd, the plume traveled south from the coast of Ibaragi through Chiba. It was believed that rain fell over a wide area in the Kanto region, and radioactive materials fell to the ground, forming relatively highly contaminated areas along the coast and southern areas of Ibaragi and around Kashiwa City in Chiba. Since the radioactive material deposited on the ground surface became a stable radiation source, its characteristics were analyzed.

\begin {figure} 
	\centering 
	\includegraphics[width=8.0cm]{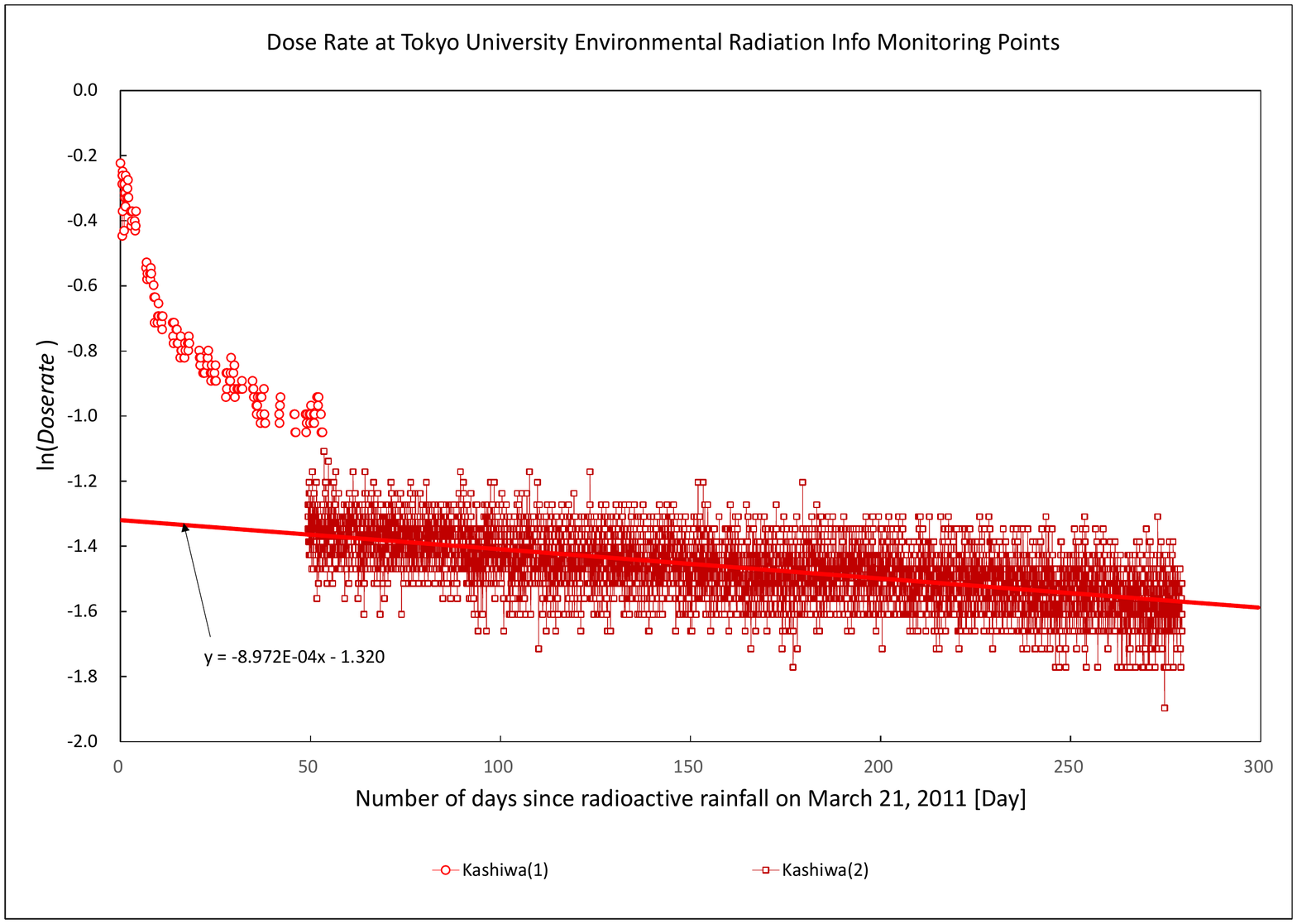} 
	\caption {The relationship between the natural logarithmic value of the dose rate at Kashiwa(1) and Kashiwa(2) ($\mu\mathrm{Sv/h}$ unit) shown in Fig. \ref{fig:dose_t_tokyouniv} and the number of days elapsed.
	}
	\label {fig:lndose_t_tokyouniv} 
\end {figure}

Figure \ref{fig:lndose_t_tokyouniv} shows the relationship between the natural logarithmic value of the dose rate at Kashiwa(1) and Kashiwa(2) ($\mu\mathrm{Sv/h}$ unit) shown in Fig. \ref{fig:dose_t_tokyouniv} and the number of days elapsed. Since the Kashiwa(2) data were almost linear, the slope of the approximate straight line gave the decay constant of 0.0008972 [1/day] using Eq. \ref{eq:lndoserate} and the half-life of $\mathrm{ln}(1/0.0008972) = 772.6$ days using Eq. \ref{eq:half-life}.

The half-life obtained from the Kashiwa(2) data was very close to the half-life of $^{134}\mathrm{C_s}$, 753.8 days\cite{JENDL}, being only 18.8 days long. So it was thought that the decrease in the Kashiwa(2) dose rate from May 10 to December 25 was mostly due to the radioactive decay of $^{134}\mathrm{C_s}$.

\begin {figure} 
	\centering 
	\includegraphics[width=8.0cm]{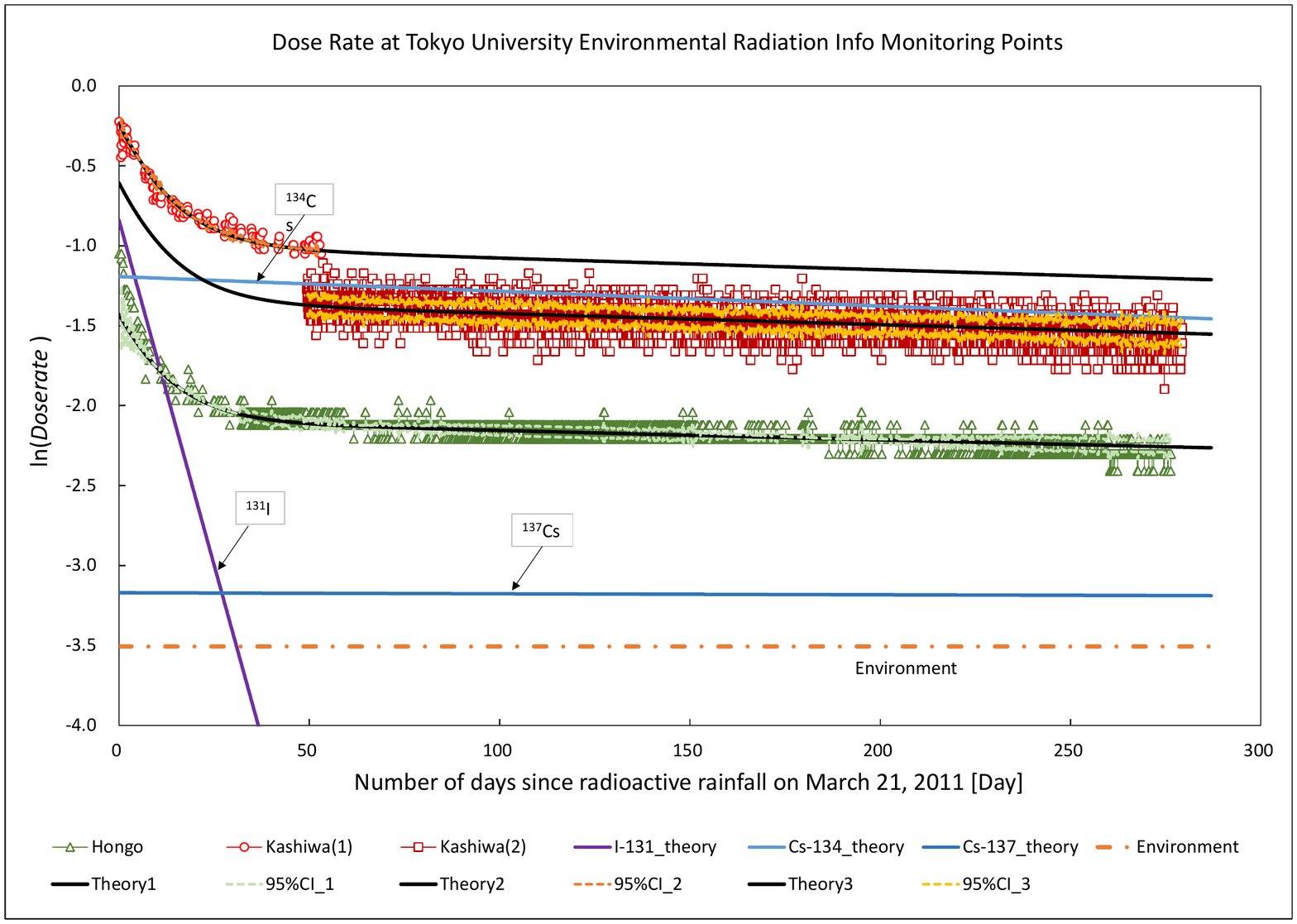} 
	\caption {The natural logarithmic value of the detected and simulated dose rate ($\mu\mathrm{Sv/h}$ unit) as a function of number of days since the radioactive rainfall on March 21, 2011. The detected values are the data at Hongo(1) and (2), and Kashiwa(1) and (2) in Fig. \ref{fig:dose_t_tokyouniv}. Black solid lines are sums of theoretical dose rates due to radioactive decay of $^{131}\mathrm{I}$ (purple line), $^{134}\mathrm{C_s}$ (blue line), and $^{137}\mathrm{C_s}$ (dark blue line) and environmental radiation dose rate (dashed line).
	}
	\label {fig:lndose_t_tokyouniv2} 
\end {figure}

Figure \ref{fig:lndose_t_tokyouniv2} shows the natural logarithmic value of the detected and simulated dose rates ($\mu\mathrm{Sv/h}$ unit) as a function of number of days since the radioactive rainfall on March 21, 2011.
The simulated dose rates due to $^{131}\mathrm{I}$, $^{134}\mathrm{C_s}$, and $^{137}\mathrm{C_s}$ and their sum were calculated using Eq. \ref{eq:doserate}. Although the values of $N_0$ in Eq. \ref{eq:doserate} were not able to be obtained precisely for each nuclide, the proportions of $^{131}\mathrm{I}$, $^{134}\mathrm{C_s}$, and $^{137}\mathrm{C_s}$ nuclei used in the calculation after adjusting to fit the data were 0.5\%, 33.0\%, and 66.5\%, respectively. 

All data measured at the University of Tokyo agreed well with the calculated curve. From this, it was thought that the significant decrease in the first 30 days was due to the radioactive decay of $^{131}\mathrm{I}$ with a half-life of about 8 days and that the slight decrease after that was mainly due to the radioactive decay of $^{134}\mathrm{C_s}$ with a half-life of about 2 years.

\subsection {Dose rate vs time at several points in Kashiwa City for 12 years}

\begin {figure} 
	\centering 
	\includegraphics[width=8.0cm]{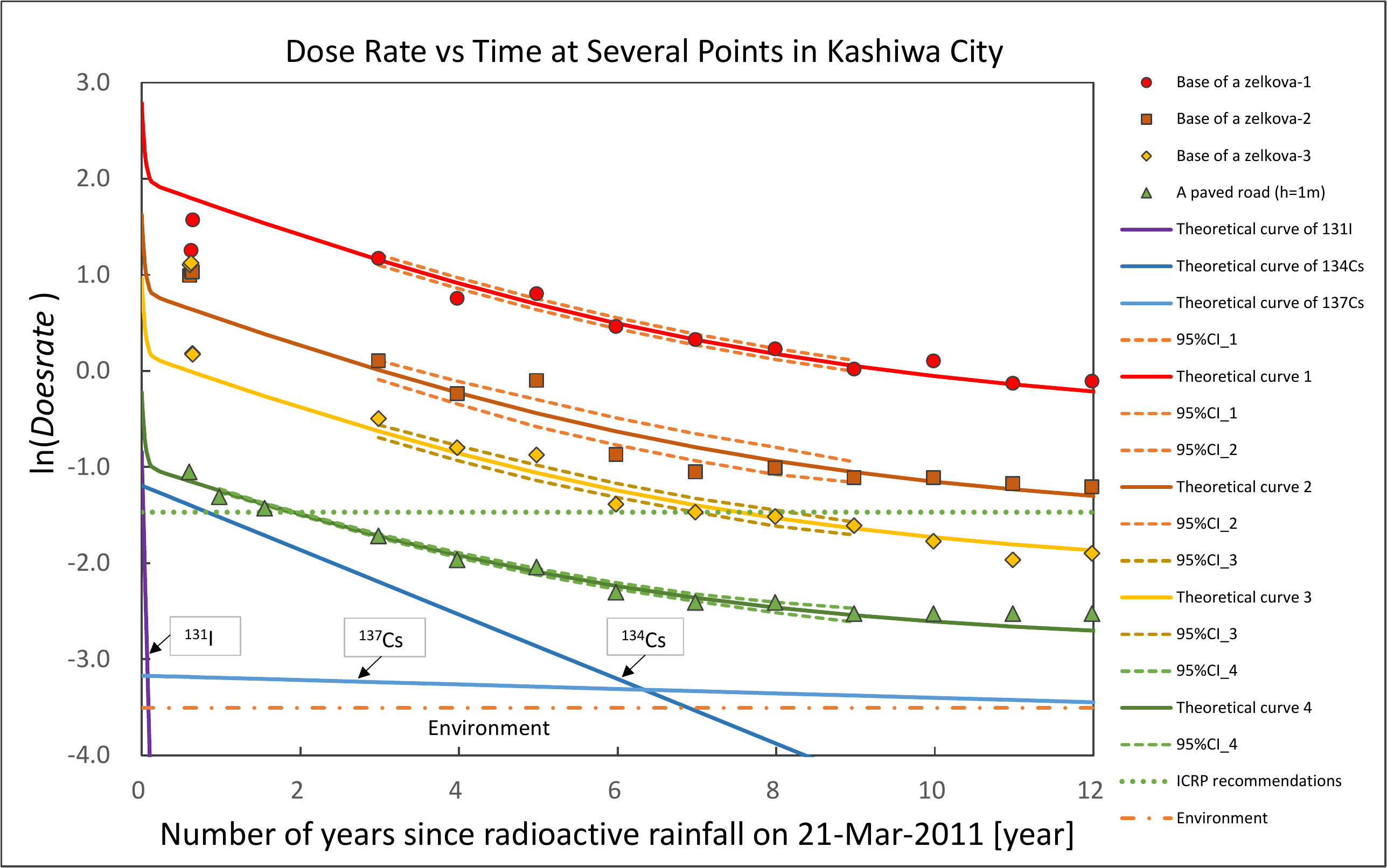} 
	\caption {The natural logarithmic value of the measured and simulated dose rate ($\mu\mathrm{Sv/h}$ unit) as a function of the number of years for 12 years since the radioactive rainfall on March 21, 2011. The circle, square, rhombus, and triangle marks indicate the measured values at the bases of three zelkova trees (point 1, 2, and 3: $h\approx\mathrm{1cm}$), and a park waling path (point 4: $h=\mathrm{1m}$), respectively. Red, ocher, yellow, and green solid lines are sums of theoretical dose rates due to radioactive decay of $^{131}\mathrm{I}$, $^{134}\mathrm{C_s}$, and $^{137}\mathrm{C_s}$ and environmental radiation dose rate.
	}
	\label {fig:lndose_t_kashiwa} 
\end {figure}

Figure \ref{fig:lndose_t_kashiwa} shows the natural logarithmic values of the measured and simulated dose rates ($\mu\mathrm{Sv/h}$ unit) as a function of the number of years since the radioactive rainfall on March 21, 2011.
The meaning of each simbol in the figure is listed in the figure caption.

 The value of $N_0$ was determined by the least squares method to fit the measured data. The environmental radiation dose rate was assumed to be $0.030 \mathrm{\mu{Sv/h}}$. The dashed line is the 95\% confidence interval obtained from moving standard deviation with a sliding window of 7.

Points 1 and 2 are the bases of zelkova trees that were not decontaminated. Measurements at these points are in close agreement with the theoretical curve for spontaneous radioactive decay. Point 3 is also the base of the zelkova tree, but the decontamination was carried out between the first measurement in March 2012 and the next measurement in October 2012, so the measured value decreased significantly. Subsequent data agree with the theoretical curve. 
The big deviation of the measured value was mainly due to the accumulation of fallen leaves and the measurement height could not be fixed.
Point 4 is a park walking path ($h=1\mathrm{m}$). Compared with the data of zelkova trees, the measured values almost agree with the theoretical curve with less deviation.

From the characteristics in Fig. \ref{fig:lndose_t_kashiwa}, it is thought that the major contribution to the radiation dose now, 12 years after the accident, is the radioactive decay of $^{137}\mathrm{C_s}$ with the  long half-life ($\approx{\mathrm{30 years}}$). People wouldn't stay at the base of the zelkova tree for long. 
On the other hand, if there is a place where rainwater collects, such as in the plaza of an apartment complex, it could be a hotspot 12 years ago and may persist for decades.
 Small children, in particular, are interested in unusual places and touch soil with their hands, so there is a possibility that soil particles with $^{137}\mathrm{C_s}$ adsorbed on them will be taken into their bodies. It is important to keep in mind the individual characteristics of places and the idiosyncrasies of human (especially children) behavior. In decontamination, to reduce the risk of radiation injury to young children, it is quite effective to identify radioactively contaminated spots in the living environment and decontaminate them intensively.

\section {Conclusions}

The present study was a report of 2011 and 2012 dose rate measurements and 12 years of fixed-point monitoring at hotspots in Kashiwa City.

A typical hotspot was a gutter and similar place where a large amount of rainwater was concentrated. The intensity distribution and time dependence of the dose rate at hotspots were measured. 

The dose rate over 12 years by fixed-point measurements agreed well with the sum of the radioactive decay characteristics of $^{131}\mathrm{I}$, $^{134}\mathrm{C_s}$, and $^{137}\mathrm{C_s}$. 

Decontamination in the living environment was conducted by sharing information between residents and the local government. 

It was effective to reduce the dose rate significantly by removing the sediment in the gutter and similar place. 
It is important to find hotspots in the living environment and decontaminate them intensively to reduce the risk of radiation injury to young children.

\section*{Appendix}

\setcounter{section}{0} 
\renewcommand{\thesection}{\Alph{section}} 
\setcounter{equation}{0} 
\renewcommand{\theequation}{\Alph{section}.\arabic{equation}}
\setcounter{figure}{0} 
\renewcommand{\thefigure}{\Alph{section}.\arabic{figure}}
\setcounter{table}{0} 
\renewcommand{\thetable}{\Alph{section}.\arabic{table}}

\section {Test of DoseRAE 2} 

DoseRAE 2 reading is updated every second. The automatic datalog can be downloaded via USB to PC. However,  the dose rate is not output via USB. Therfore, the relationship between the dose rate and time was obtained by taking a video of the displayed value. In Kashiwa city, radiation sources ($\approx{0.1 \mathrm{{\mu}Sv/h}}$ to $12 \mathrm{{\mu}Sv/h}$) were nearby. This allowed us to test the dynamic response of the dosimeter. The test was carried out in January 2012\cite{DoseRAE2test}.

\subsection {Starting characteristics}

\begin {figure} [p] 
	\centering 
	\includegraphics[width=8.0cm]{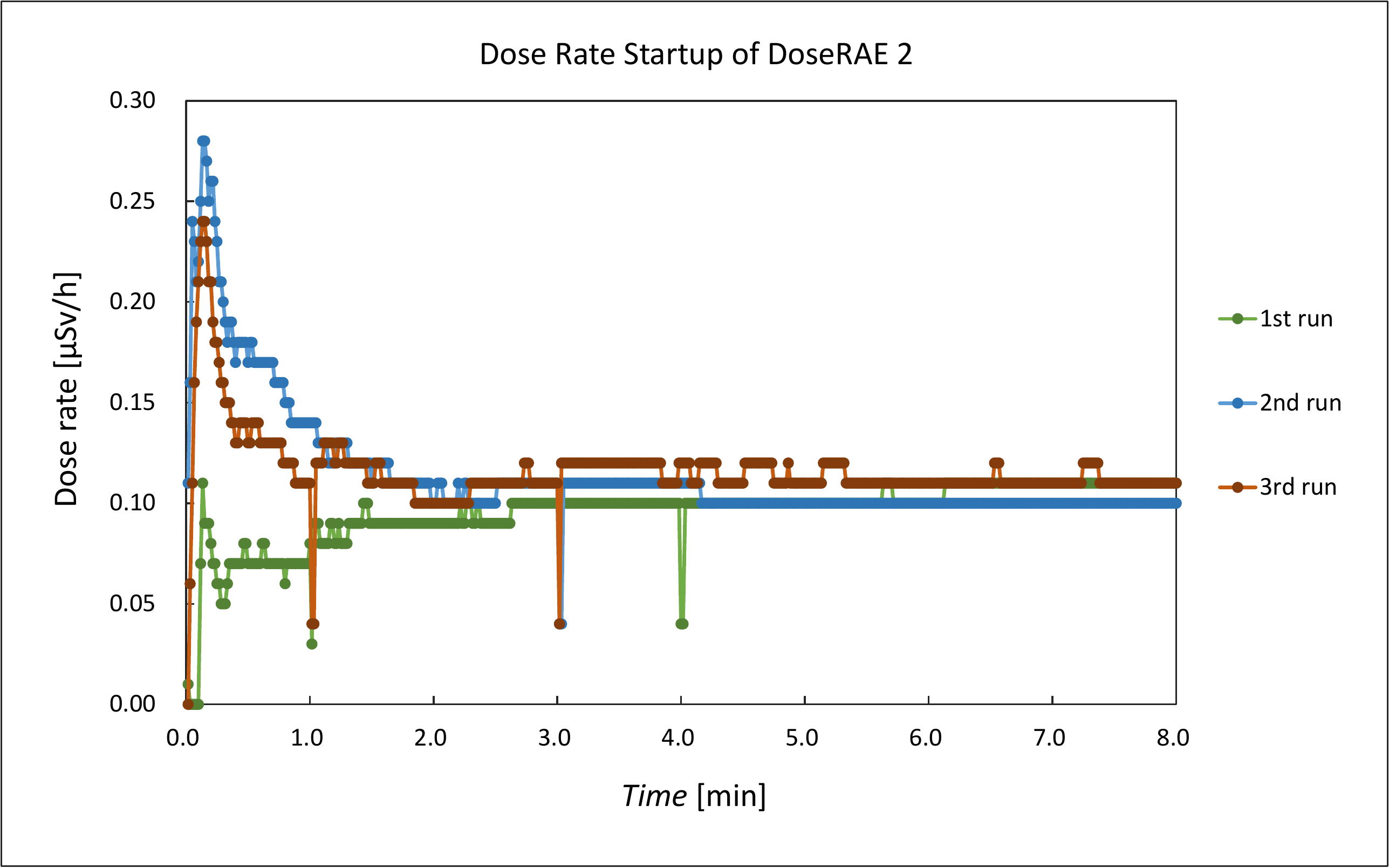} 
	\caption {The starting characteristics of dose rate vs time measured three times.
	}
	\label {fig:startup} 
\end {figure}

Figure \ref{fig:startup}  shows the starting characteristics of dose rate vs time measured three times (1st run, 2nd run, 3rd run). It overshooted after 3 to 7 seconds of dead time and displayed a steady value after about 2 minutes. After that, the value showed an instantaneous drop and a gentle change, and stabilized in about 6 minutes. The steady-state value at the measurement location was between $0.09 \mathrm{{\mu}Sv/h}$ and $0.11 \mathrm{{\mu}Sv/h}$ and the display was stable. 

\subsection {Dose rate vs time characteristics in a region around $0.1 \mathrm{{\mu}Sv/h}$}

\begin {figure} [p] 
	\centering 
	\includegraphics[width=8.0cm]{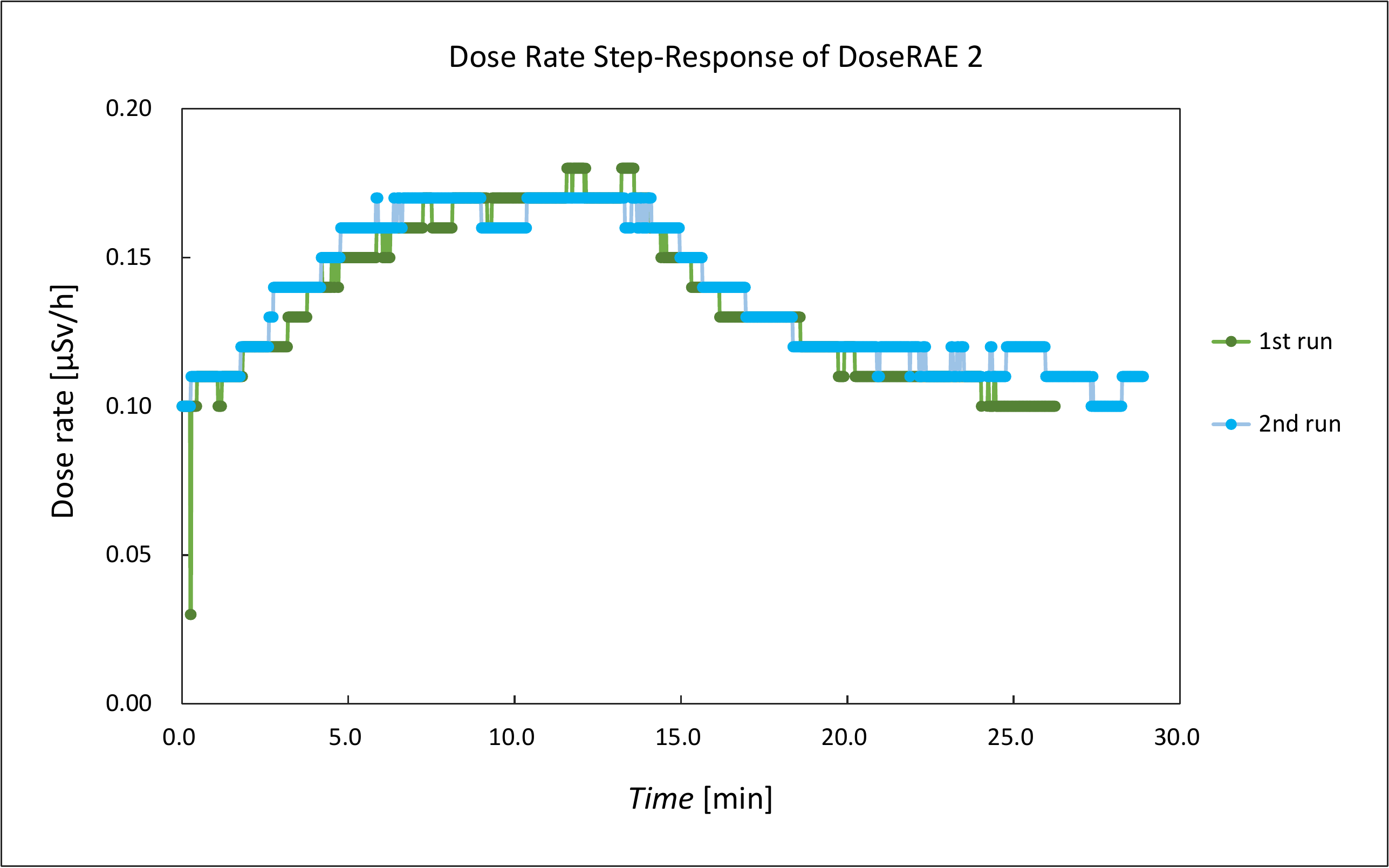} 
	\caption {The dose rate vs time characteristics in a low dose rate ($\approx\mathrm{0.1{\mu}Sv/h}$) region.
	}
	\label {fig:Dose_t_step02microSvh} 
\end {figure}

Figure \ref{fig:Dose_t_step02microSvh} shows the dose rate vs time characteristics in a low dose rate ($\approx\mathrm{0.1{\mu}Sv/h}$) region.
A weakly radioactive radium ($\mathrm{R_a}$) ball was placed on the DoseRAE 2 sensor to measure the increase in dose rate. After reaching a steady-state value, the ball was removed and the decrease was measured.
The dose rate gradually increased and decreased and reached a steady value in about 10 minutes.
The data were reproducible and read a meaningful change of $0.01 \mathrm{{\mu}Sv/h}$.

\subsection {Dose rate vs time characteristics in a region around $1 \mathrm{{\mu}Sv/h}$}

\begin {figure} 
	\centering 
	\includegraphics[width=8.0cm]{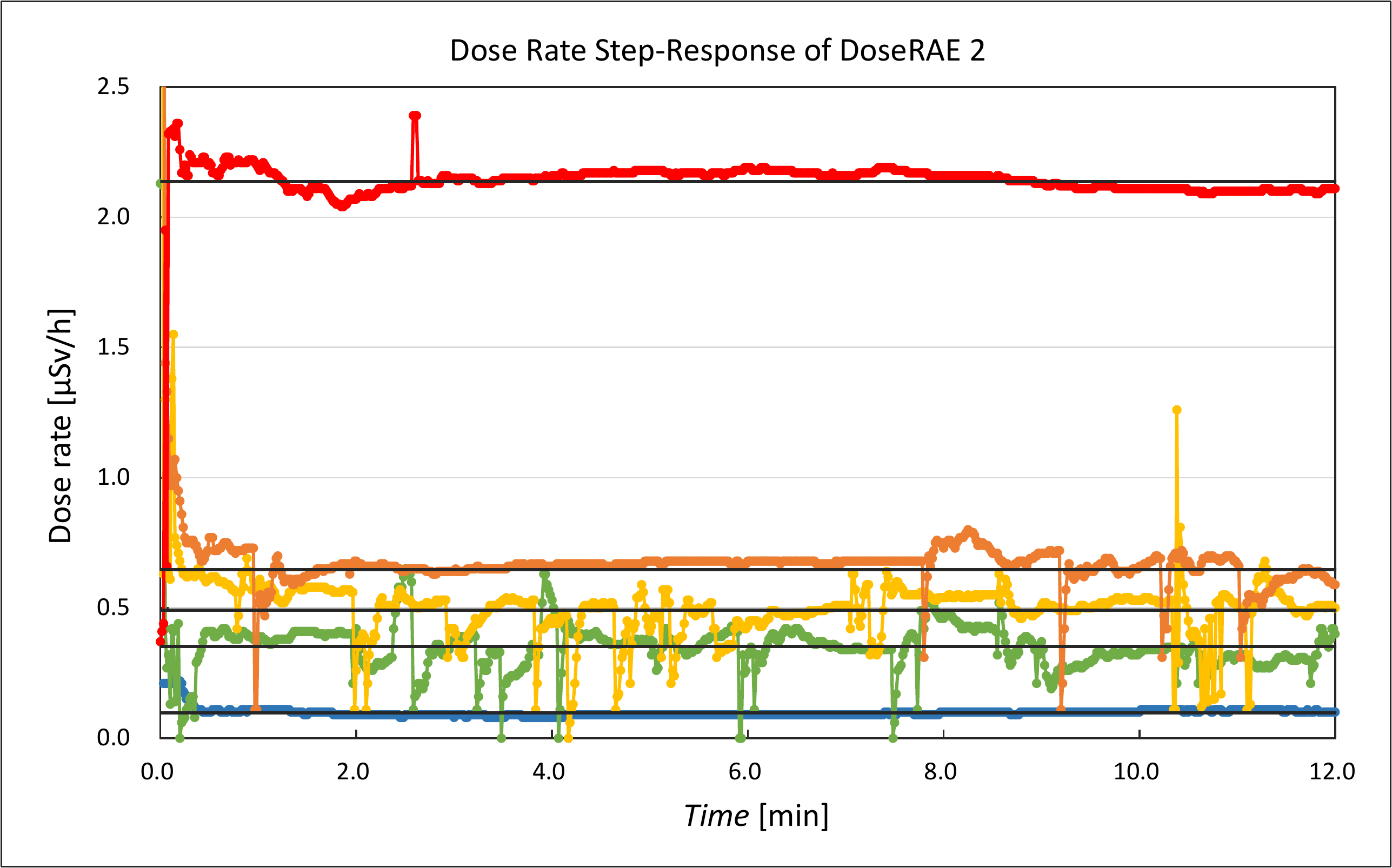} 
	\caption {The time dependence of dose rates below about $2 \mathrm{{\mu}Sv/h}$.
	}
	\label {fig:Dose_t_step01to2microSvh} 
\end {figure}

Figure \ref{fig:Dose_t_step01to2microSvh} shows the time dependence of dose rates below about $2 \mathrm{{\mu}Sv/h}$. The measurement location was the park parking lot. The characteristics showed a steady value after about 3 minutes, so the figure shows the average values from 3 minutes to 12 minutes with thin solid lines. The average values of the five characteristics in Fig. \ref{fig:Dose_t_step01to2microSvh} are 0.10, 0.35, 0.48, 0.66, and $2.14 \mathrm{{\mu}Sv/h}$, respectively.

Table \ref{table:Dose_t_step01to2microSvh} shows the maximum and minimum values [\%] of deviations from the average value (from 3 to 12 minutes) for each dose rate characteristic.

When the dose rate is about $0.1 \mathrm{{\mu}Sv/h}$ the time for steady state $\approx\mathrm{10 min}$ as shown in Fig. \ref{fig:Dose_t_step02microSvh} and the deviations from the average value were within about $\pm{15 \%}$ as shown in Table \ref{table:Dose_t_step01to2microSvh}.

\begin{table} [h] 
  \caption {The maximum and minimum values [\%] of deviations from the average value (from 3 to 12 minutes) for each dose rate characteristic.\\} 
  \label {table:Dose_t_step01to2microSvh}
  \centering 
  \begin {tabular} {crrrrr} 
	\hline 
	Ave. & \multicolumn{1}{c}{0.10} & \multicolumn{1}{c}{0.35} & \multicolumn{1}{c}{0.48} & \multicolumn{1}{c}{0.66} & \multicolumn{1}{c}{2.14} \\ 
	\hline \hline 
	Max & 15.3 & 82.5 & 162.1 & 20.7 & 2.1 \\ 
        Min & -16.1 & -100 & -100 & -83.4 & -2.6 \\
	\hline
  \end {tabular} 
\end {table}

\subsection {Dose rate vs time characteristics in a region around $0.5 \mathrm{{\mu}Sv/h}$}

The characteristics with average values of 0.35, 0.48, and $0.66 \mathrm{{\mu}Sv/h}$ had remarkably large fluctuations as shown in Table \ref{table:Dose_t_step01to2microSvh}. It was necessary to test whether reliable values could be obtained from fluctuating data. On the lawn in the park where there was little disturbance, the displayed values were video-recorded for about 10 minutes at different heights of 1, 10, 30, 50, 75, 100, and $\mathrm{120 cm}$.

\begin {figure} 
	\centering 
	\includegraphics[width=8.0cm]{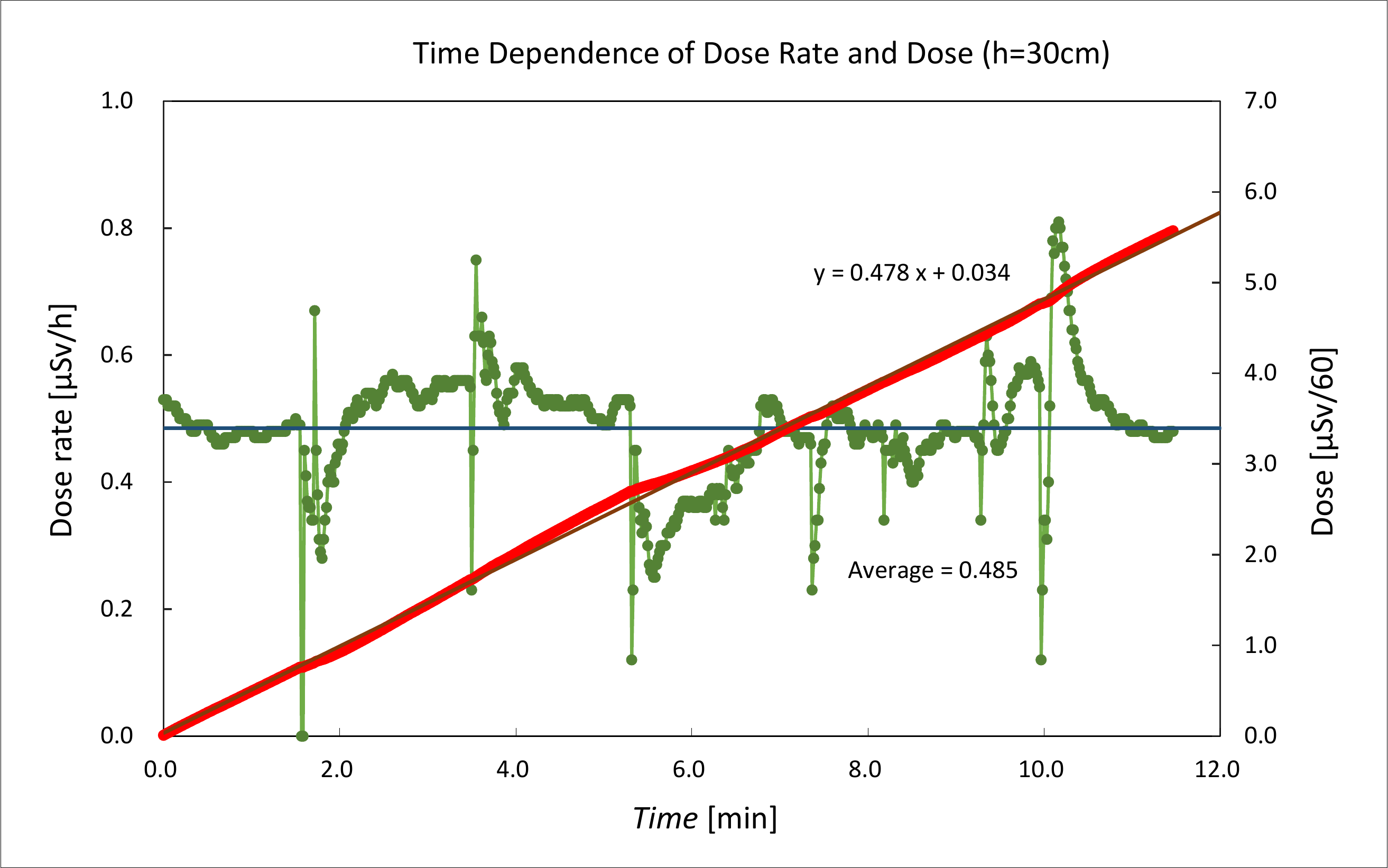} 
	\caption {The time dependence of the dose rate (dark green marker) and its average value (black horizontal line), the cumulative dose (red marker) and its approximation line (dark brown straight line).
	}
	\label {fig:Dose_drate_t_05h30} 
\end {figure}

All measurements showed large fluctuations when the dose rates were around $0.5 \mathrm{{\mu}Sv/h}$. Figure \ref{fig:Dose_drate_t_05h30} shows the time [min] dependence of the dose rate [$\mathrm{{\mu}Sv/h}$] (dark green marker) and its average value (black horizontal line), the cumulative dose [$\mathrm{{\mu}Sv/60}$] (red marker) and its approximation line (dark brown straight line) at a height of $\mathrm{30 cm}$. The average dose rate was $0.485 \mathrm{{\mu}Sv/h}$. The slope of the approximation line of the cumulative dose was $ 0.478 \mathrm{{\mu}Sv/h}$, which was almost equal to the average value.

\begin {figure} 
	\centering 
	\includegraphics[width=8.0cm]{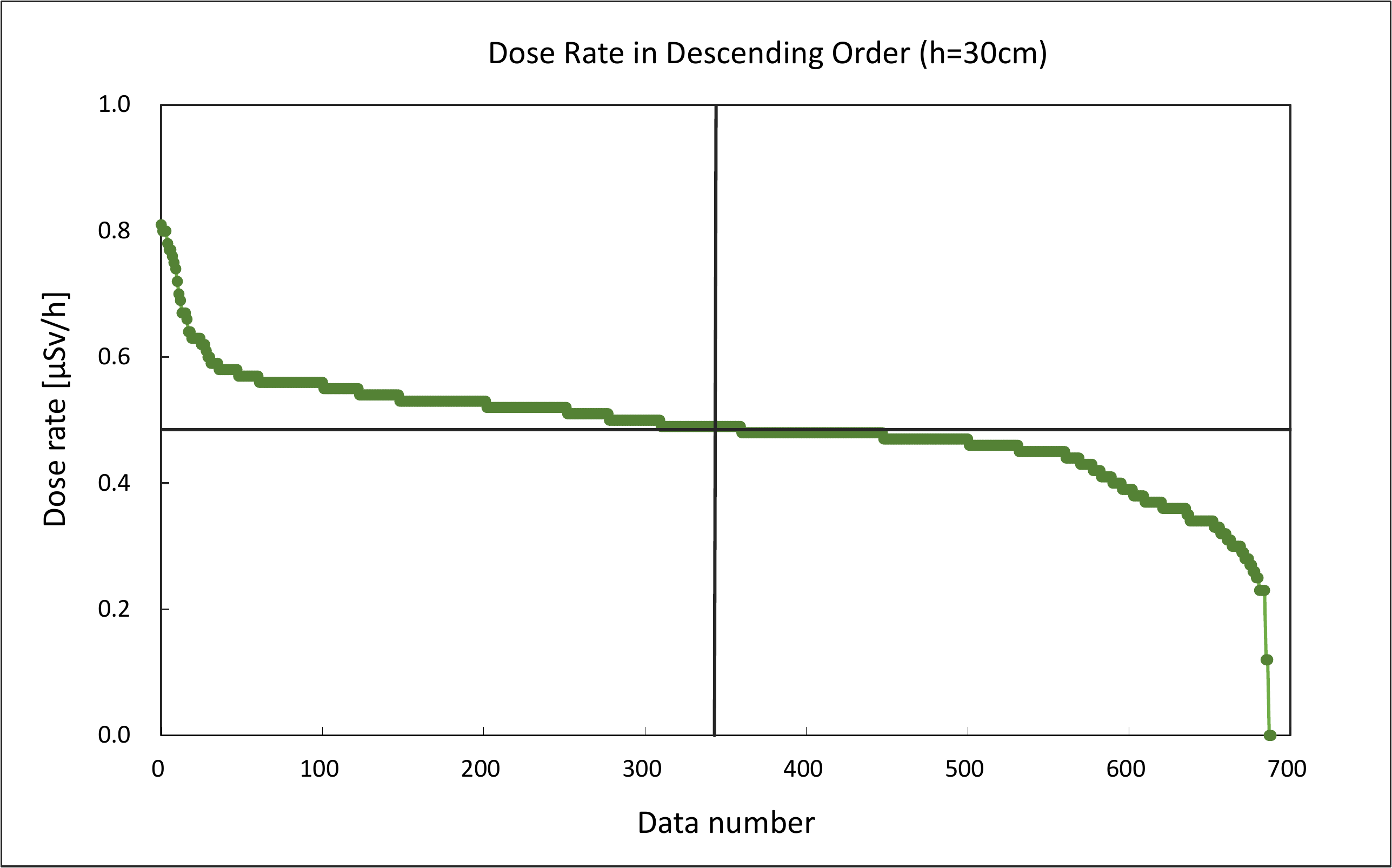} 
	\caption {The relationship between the dose rate in descending order and the data number.
	}
	\label {fig:Dosedescending_n_05h30} 
\end {figure}

The dose rates of 600 data in Fig. \ref{fig:Dose_drate_t_05h30} were arranged in descending order and given data numbers from 0 to 599. Figure \ref{fig:Dosedescending_n_05h30} shows the relationship between the dose rate and the data number. The dose rate of data number 300 is the median. The dose rates were almost symmetrical about the median. Therefore, the average dose rate (horizontal solid line) and the median were close to each other.

Summarizing the above results, Table \ref{table:Mean_Median_Slope} shows (1) the average value, (2) the median value, and (3) the cumulative approximate linear gradient for each measured height. The values of (1), (2), and (3) were close to each other, despite the large fluctuation in the measured values. Deviations from the mean values of (1), (2), and (3) fell between 2.6 \% and -2.3\%. When the metastable measured values were read by eye, the deviations were within $\pm{30\%}$. However, if five metastable values were read in about 10 minutes and their average was taken, it would be within $\pm{3\%}$.

\begin{table} [h] 
  \caption {The average value, the median value, and the cumulative approximate linear gradient [$\mathrm{{\mu}Sv/h}$] measured at different heights above ground [cm].\\} 
  \label {table:Mean_Median_Slope}
  \centering 
  \begin {tabular} {crrrr} 
	\hline 
	Height & \multicolumn{1}{c}{1} & \multicolumn{1}{c}{10} & \multicolumn{1}{c}{30} & \multicolumn{1}{c}{50} \\ 
	\hline \hline 
	Average & 0.504 & 0.511 & 0.485 & 0.473 \\ 
        Median & 0.52 & 0.51 & 0.49 & 0.49  \\ 
	Gradient & 0.497 & 0.511 & 0.478 & 0.486 \\ 
	\hline
	\\
	\hline 
	Height & \multicolumn{1}{c}{50} & \multicolumn{1}{c}{75} & \multicolumn{1}{c}{100} & \multicolumn{1}{c}{120} \\ 
	\hline \hline 
	Average & 0.473 & 0.481 & 0.478 & 0.486 \\ 
        Median & 0.49 & 0.48 & 0.49 & 0.50 \\ 
	Gradient & 0.486 & 0.479 & 0.472 & 0.477 \\ 
	\hline
  \end {tabular} 
\end {table}

It was presumed that the cause of the large fluctuation was the division of the accumulated dose increment by a short time increment in the region where the output of the sensor was small.

\subsection {Dose rate vs time characteristics in a region around $10 \mathrm{{\mu}Sv/h}$}

\begin {figure} 
	\centering 
	\includegraphics[width=8.0cm]{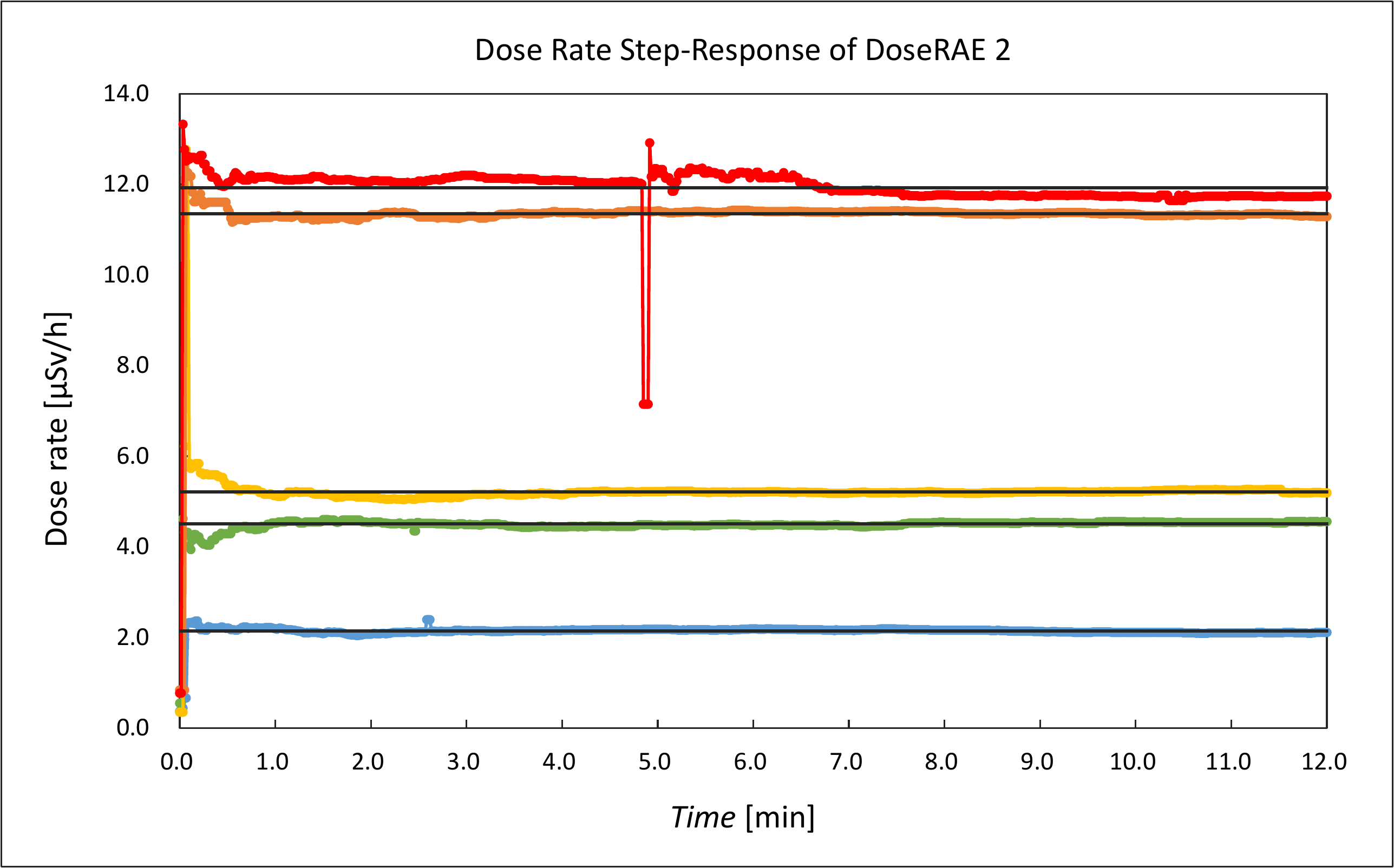} 
	\caption {The time dependence of dose rates from $2 \mathrm{{\mu}Sv/h}$ to about $12 \mathrm{{\mu}Sv/h}$.
	}
	\label {fig:Dose_t_step2to12microSvh} 
\end {figure}

Figure \ref{fig:Dose_t_step2to12microSvh} shows the time dependence of dose rates from $2 \mathrm{{\mu}Sv/h}$ to about $12 \mathrm{{\mu}Sv/h}$. The measurement location was the park parking lot. The characteristics showed a steady value after about 3 minutes, so the figure shows the average values from 3 minutes to 12 minutes with thin solid lines. The average values of the five characteristics in Fig. \ref{fig:Dose_t_step2to12microSvh} are 2.14, 4.50, 5.21, 11.37, and $11.92 \mathrm{{\mu}Sv/h}$, respectively.

Table \ref{table:Dose_t_step2to12microSvh} shows the maximum and minimum values [\%] of deviations from the average value (from 3 to 12 minutes) for each dose rate characteristic.

\begin{table} [h] 
  \caption {The maximum and minimum values [\%] of deviations from the average value (from 3 to 12 minutes) for each dose rate characteristic.\\} 
  \label {table:Dose_t_step2to12microSvh}
  \centering 
  \begin {tabular} {crrrrr} 
	\hline 
	Ave. & \multicolumn{1}{c}{2.14} & \multicolumn{1}{c}{4.50} & \multicolumn{1}{c}{5.21} & \multicolumn{1}{c}{11.37} & \multicolumn{1}{c}{11.92} \\ 
	\hline \hline 
	Max & 2.1 & 1.2 & 1.2 & 0.6 & 3.7 \\ 
        Min & -2.6 & -1.7 & -1.5 & -0.9 & -2.2 \\
	\hline
  \end {tabular} 
\end {table}

From Table \ref{table:Dose_t_step2to12microSvh}, when the dose rate is about $2 \mathrm{{\mu}Sv/h}$ or more, the measured values after about 3 minutes had deviations within $\pm{3 \%}$ from the average value.



\begin{thebibliography}{14} 
	\bibitem{INES} The International Nuclear and Radiological Event Scale User's Manual, 2008 Edition. \\
		\url{https://www-pub.iaea.org/MTCD/Publications/PDF/INES2013web.pdf}

	\bibitem{NHKscicul2023} NHK Science and culture, 2023.04.10. \\
		\url{https://www3.nhk.or.jp/news/special/sci_cul/2023/04/story/nuclear20230407/}

	\bibitem{NHKscicul2019} NHK Science and culture, 2019.02.14. \\
		\url{https://www3.nhk.or.jp/news/special/sci_cul/2019/02/story/special_190214/}

	\bibitem{Report_NERH} Report of the Nuclear Emergency Response Headquarters,  April 8, 2011. \\
		\url{https://www.kantei.go.jp/saigai/pdf/201104081330genpatsu.pdf}

	\bibitem{Report_IAEA} Report submitted to the IAEA by the Government of Japan, June 2011. \\
		\url{https://www.kantei.go.jp/jp/topics/2011/iaea_houkokusho.html}

	\bibitem{Aircraft_monitoring} Airborne monitoring by the Ministry of Education, Culture, Sports, Science and Technology. \\
		\url{https://www.city.kawaguchi.lg.jp/material/files/group/14/1910_092917_1.pdf}

	\bibitem{Guideline-localcontami} The Guide line for the local contamination by the radioactive materials, March 2012, Ministry of environment. \\
		\url{https://www.env.go.jp/content/900483790.pdf}

	\bibitem{Kashiwa-contami_survay} The detailed survay of radioactive contamination in Kashiwa city, Dec. 28, 2011, Ministry of environment. \\
	\url{https://www.env.go.jp/press/files/jp/18940.pdf}

	\bibitem{ICRP} ICRP Publication 103, ICRP, 2007. \\
		\url{https://www.icrp.org/publication.asp?id=ICRP%20Publication%20103}

	\bibitem{DoseRAE2test} Y. Takase, Dynamic Response Test of Dosimeter DoseRAE 2. \\
		\url{https://drive.google.com/file/d/1-NuSBttX8DqUBTuBlLmEcFW7VCx4wI-u/view?usp=sharing}

	\bibitem{Yamazawa-Hirao} H. Yamazawa and S. Hirao, Environmental influence through air by the Fukushima nuclear disaster, Proceedings of the Japanese nuclear Association, Vol. 53, No. 7, 2011.

	\bibitem{TokyoUnivERI} The environmental radiation information of the University of Tokyo. \\
		\url{https://www.u-tokyo.ac.jp/ja/administration/erc/}

	\bibitem{Rutherford} E. Rutherford and F. Soddy, Philosophical Magazine 4, 370-96 (1902). \\
		\url{https://web.lemoyne.edu/~giunta/ruthsod.html}

	\bibitem{JENDL} Japan Atomic Energy Agency, Tables of Nuclear Data. \\
		\url{https://wwwndc.jaea.go.jp/NuC/index.html}

\end{thebibliography}
\end{document}